\documentclass[usenatbib]{mn2e}
\usepackage{epsfig}
\def\gtwid{\mathrel{\raise.3ex\hbox{$>$\kern-.75em\lower1ex\hbox{$\sim$}}}}
\def\ltwid{\mathrel{\raise.3ex\hbox{$<$\kern-.75em\lower1ex\hbox{$\sim$}}}}
\def\\{\hfil\break}
\def\ie{{\it i.e.\ }}

\def\lesssim{\mathrel{\hbox{\rlap{\hbox{\lower4pt\hbox{$\sim$}}}\hbox{$<$}}}}
\def\gtrsim{\mathrel{\hbox{\rlap{\hbox{\lower4pt\hbox{$\sim$}}}\hbox{$>$}}}}

\def\arcdeg{\hbox{$^\circ$}}

\newcommand{\mamo}[1]{\mbox{$#1$}}
\newcommand{\unit}[1]{\ifmmode \:\mbox{\rm #1}\else \mbox{#1}\fi}
\newcommand{\sbr}[1]{_{\rm #1}}
\newcommand{\mone}{\mamo{^{-1}}}

\newcommand{\kms}{\unit{km~s\mone}}

\newcommand{\mpc}{\unit{Mpc}}

\newcommand{\hmpc}{\mamo{h\mone}\mpc}

\newcommand{\lb}[2]{\mamo{l = #1\arcdeg}, \mamo{b = #2\arcdeg}}

\newcommand{\lberr}[4]{\mamo{l = #1\pm#2\arcdeg}, \mamo{b = #3\pm#4\arcdeg}}

\newcommand{\Om}{\mamo{\Omega\sbr{m}}}

\begin{document}

\title[Large Cosmic Flows on Scales of 100\hmpc]
{Consistently Large Cosmic Flows on Scales of 100\hmpc: a Challenge for the Standard $\Lambda$CDM Cosmology}
\vskip 0.5cm
\author[Watkins, Feldman \& Hudson]{Richard Watkins$^{\dagger,1}$, Hume A. Feldman$^{\star,2}$ \& Michael J. Hudson$^{\ddagger,3}$\\
$^\dagger$Department of Physics, Willamette University, Salem, OR 97301, USA.\\
$^\star$Department of Physics \& Astronomy, University of Kansas, Lawrence, KS 66045, USA.\\
$^\ddagger$Department of Physics and Astronomy, University of Waterloo, Waterloo, ON N2L 3G1, Canada.\\
emails: $^1$rwatkins@willamette.edu;\, $^2$feldman@ku.edu;\, $^3$mjhudson@uwaterloo.ca}

\maketitle

\begin{abstract}

The bulk flow, \ie the dipole moment of the peculiar velocity field, is a sensitive probe of matter density fluctuations on very large scales.   However, the peculiar velocity surveys for which the bulk flow has been calculated have non-uniform spatial distributions of tracers, so that the bulk flow estimated does not correspond to that of a simple volume such as a sphere.  Thus bulk flow estimates are generally not strictly comparable between surveys, even those whose effective depths are similar.   In addition, the sparseness of typical surveys can lead to aliasing of small scale power into what is meant to be a probe of the largest scales.   Here we introduce a new method of calculating bulk flow moments where velocities are weighted to give an optimal estimate of the bulk flow of an idealized survey, with the variance of the difference between the estimate and the actual flow being minimized.    These ``minimum variance'' estimates can be designed to estimate the bulk flow on a particular scale with minimal sensitivity to small scale power, and are comparable between surveys.   
We compile all major peculiar velocity surveys and apply this new method to them.  We find that most surveys we studied are highly consistent with each other. Taken together the data suggest that the bulk flow within a Gaussian window of radius 50 \hmpc\ is $407\pm81$ \kms\ toward  \lb{287\arcdeg\pm9}{8\arcdeg\pm6}.  The large-scale bulk motion is consistent with predictions from the local density field. This indicates that there are significant density fluctuations on very large scales.  A flow of this amplitude on such a large scale is not expected in the WMAP5-normalized $\Lambda$CDM cosmology, for which the predicted one-dimensional r.m.s.\ velocity is $\sim 110$ \kms. The large amplitude of the observed bulk flow favors the upper values of the WMAP5 $\Om h^2$-$\sigma_8$ error-ellipse, but even the point at the top of the WMAP5 95\% confidence ellipse predicts a bulk flow which is too low compared to that observed at $>98$\% confidence level.
\end{abstract}

\noindent{\it Subject headings}: cosmology: distance scales -- cosmology: large scale structure of the universe -- cosmology: observation -- cosmology: theory -- galaxies: kinematics and dynamics -- galaxies: statistics

\section{Introduction}
\label{sec:intro}

A long-standing question in cosmography is the origin of the $\sim 600$ km/s peculiar velocity of the Local Group (LG) with respect to the Cosmic Microwave Background (CMB).  The motion of the LG with respect to the ``Local Sheet'' in which it is embedded is only $\sim 60$ km/s \citep{TulShaKar08}, thus most of the LG's motion is due to structures on scales larger than the Local Sheet, i.e.\ beyond  5 \hmpc\ (where $h$ is the Hubble constant in units of 100 km s$^{-1}$ Mpc$^{-1}$). In the gravitational instability paradigm \citep{FelFriFry01,ScoFelFri01,VerHeaPer02},
this motion is due to the gravity of structures on larger scales. For a galaxy at position {\bf{r}}, the peculiar velocity {\bf{v}} is given by \citep{PPC}
\begin{equation}
{\bf{v}}\left( {\bf{r}} \right) =
\frac{\Omega\sbr{m}^{0.55}}{4\pi }\int d^3 {\bf{r}}^{\prime
} \delta\sbr{m}\left( \mathbf{r}^{\prime }\right) \frac{\left( \mathbf{r}^{\prime
}-\mathbf{r}\right) }{\left| \mathbf{r}^{\prime }-\mathbf{r}\right| ^3}.
\label{eq:peculiar}
\end{equation}
where $\delta\sbr{m} \left( \mathbf{r}\right) = ({\rho
  -\overline{\rho}})/{\overline{\rho}}$, and $\overline{\rho}$ is the average density of the Universe,  $\Om$ is the matter density parameter, and we have used 0.55 instead of 0.6 for the power of $\Omega_m$ in the pre-factor to improve accuracy for models with dark energy \citep{Lin05}. 

The issue of the LG's motion and that of other nearby galaxies has important cosmological and cosmographical implications. Specifically, as shown by Eq. \ref{eq:peculiar}, the peculiar velocities of individual galaxies are sensitive to the matter power spectrum over a wide range of scales. Indeed, apart from the Integrated Sachs-Wolfe (ISW) effect \citep{SacWol67}, peculiar velocities are the only probe of the matter density fluctuations on scales of $\sim 100 \hmpc$ and clearly the only dynamical probe in the low-redshift universe. A given power spectrum predicts the r.m.s.\ of the components of a galaxy's peculiar motion. For models with more power, i.e. a higher normalization, one predicts a larger r.m.s.\ velocity. 

For a single galaxy, the contributions to its motion arise from a range of scales: from the local ($\sim$ 5 \hmpc) to the very large ($\ga 100 \hmpc$) scales. One may reduce the effects of small scale density fluctuations by studying the peculiar velocity of a larger volume using a sample of peculiar velocity tracers such as galaxies, clusters, or Type Ia supernovae. Beginning with the work of \cite{RubRobTho76}, a number of such surveys have been undertaken over the last couple of decades 
\citep{
DreFabBur87,
LynFabBur88,
AarBotCor89,
Wil90,
Cou92,
HanMou92,
MatForBuc92b,
LauPos94,
HudLucSmi97,
WilCouFab97,
GioHayFre98,
HudSmiLuc99,
DalGioHay99,
DalGioHay99b,
Wil99b,
CouWilStr00,
ColSagBur01,
HudSmiLuc04,
HauHanTho06,
SprMasHay08}.  

The simplest statistic that can be derived from a sample of peculiar velocities is the dipole moment of the sample, also known as its bulk flow. It was quickly realized that the bulk flow was closely related to the amplitude of fluctuations on large scales, and could be used to test cosmological models \citep{CluPee81, VitJusDav86}. At face value, however, the surveys cited above yield apparently conflicting results: the measured bulk flow ranges from 0 to $\sim 1000$ km/s.  Note, however, that many of the above-mentioned surveys are sparsely-sampled, and that while authors quote the bulk flow of the sample, this sample bulk flow is often mis-interpreted as the coherent bulk flow of the whole volume occupied by the survey. 

The issue of sparse sampling, small-scale aliasing and their effects on statistics such as the bulk flow were first analyzed by \cite{Kai88}, and later \citet{WatFel95} and others \citep{pairwise00,Hud03,pairwise03,SarFelWat07,WatFel07,FelWat08}. These studies addressed the issue of comparing sparse surveys both to each other (to check for consistency between different sparse surveys) and the comparison of sparse peculiar velocity samples with expectations from cosmological models. One lesson from this work is that both sparse sampling and aliasing present an important effect that must be accounted for in interpreting the results, particularly those from sparse surveys such as clusters or SNe. 

Bulk flow estimates are essentially weighted averages of the individual velocities in a survey.   Previous work has focused on a weighting scheme that produces a maximum likelihood estimate (MLE) of the bulk flow of a survey, an estimate that minimizes the uncertainties due to measurement noise but does not make any correction for the survey geometry .  Thus the MLE bulk flow is obviously dependent on a given survey's particular geometry and statistical properties.  In this paper, we instead address the question of how peculiar velocity data can be used to estimate a more general statistic: the bulk flow of an ideal, densely-sampled survey with a given depth.   Our approach will be to calculate \emph{optimal} weights which produce the best possible estimate of this statistic.  An approach related to this question is that of Zaroubi and collaborators, who used Wiener filtering \citep{ZarHofDek99} and variants \citep{Zar02} to reconstruct the matter density field directly from peculiar velocities.  That work, however, was focussed more on the mapping of the density field and the measurement of $\beta = f(\Om)/b$,  where $b$ is the bias parameter, than on the bulk flow \cite[but see][]{HofEldZar01}. In this paper, our aim is somewhat different: to construct dipole moments that probe the largest scales.

The goal of this paper is to make the cleanest measurement of the large-scale bulk flow using the best peculiar velocity data available. We discuss the peculiar velocity surveys used in this analysis in Section~\ref{sec:surveys}.  In Section~\ref{sec:mom},  we describe the construction of the velocity moments,  the power spectrum model,  and the optimal weighting scheme used to estimate bulk flow components free of small scale noise. In  Section~\ref{sec:res} we apply these optimal weights to the data. In Section~\ref{sec:consist}, we assess whether the optimally-weighted bulk flow results from different surveys are mutually consistent. In Section~\ref{sec:implic}, we discuss the cosmographic implications of our results. In Section~\ref{sec:compare}, we compare the measured bulk flow with expectations from cosmological models. We discuss our resuts in Section~\ref{sec:discuss} and conclude in Section~\ref{sec:conc}. 

\section{Data}
\label{sec:surveys}

Here we analyze all of the recent peculiar velocity surveys. The datasets occasionally have outliers, and so it is necessary to remove them. Simply removing outliers with large CMB velocities might bias the resulting flow. Instead, we use predictions from the IRAS-PSCz density field to identify outliers, according to the following procedure. First, we compare the observed peculiar velocity with the predicted peculiar velocity from \cite{HudSmiLuc04}, adopting the parameters of the B05 ($\beta = 0.5$) flow model used by \cite{NeiHudCon07}. This model allows for a small external bulk flow arising from large scales, but provides a better predicted peculiar velocity within the PSCz volume, \ie within a distance of 200 \hmpc. Peculiar velocities that deviate by more than 3.5 $\sigma$ are rejected, where the uncertainty includes the distance error and a thermal component of 150 \kms.   For each sample, we also quote a characteristic depth defined as the mean weighted distance, where the weight is the inverse square of the peculiar velocity error. 

The final samples are as follows, listed in order of characteristic depth from nearest to most distant.

\begin{itemize}

\item SBF: the surface brightness fluctuation survey of \cite{TonDreBla01}. We use the distances from their Table 1, except where SBF galaxies are identified as group members, in which case we use the group peculiar velocities (their Table 4).  After rejection of the outliers N4616, N4709 and ESO 323-034, there are 69 field and 23 groups, with a characteristic depth of  17 \hmpc.

\item ENEAR: a survey of Fundamental Plane (FP) distances to nearby early-type galaxies \citep{daCBerAlo00, BerAlodaC02b, WegBerWil03} . After the exclusion of 4 outliers, there are distances to 698 field galaxies or groups (Bernardi, priv. comm). For single galaxies, the typical distance error is $\sim 20\%$. The characteristic depth of the sample is 29 \hmpc. Note that unlike other samples considered here, these data are not corrected for inhomogeneous Malmquist bias \citep{Hud94}.

\item SN are 103 Type Ia supernovae distances from the compilation of \cite{TonSchBar03}, limited to a distance of 150 \hmpc. SN distances are typically precise to 8\%. The characteristic depth of the survey is 32 \hmpc.

\item SFI++  \citep{SprMasHay08}, based on the Tully-Fisher (TF) relation, is the largest and densest peculiar velocity survey considered here. After rejection of 38 (1.4\%) field and 10 (1.3\%) group outliers, our sample consist of 2675 field galaxies and 726 groups. For some analyses, we split this large sample into a field (SFI++$_F$) and group (SFI++$_G$) subsamples. The characteristic depth of SFI++ is 34 \hmpc.

\item SC \citep{GioHaySal98, DalGioHay99} is a TF-based survey of spiral galaxies in 70 clusters within 200 \hmpc. The characteristic depth of the combined sample is 57 \hmpc. 

\item SMAC \citep{HudSmiLuc99, HudSmiLuc04} is an all-sky  Fundamental Plane (FP) survey of 56 clusters. The characteristic depth of the survey is 65 \hmpc.

\item LP \citep{LauPos94,PosLau95} is a survey based on using brightest cluster galaxies as distance indicators. The survey consists of BCGs in 119 Abell clusters across the whole sky within a distance of 150 \hmpc. Here we obtain peculiar velocities using the methodology, but not the X-ray correction, of \cite{HudEbe97}, which makes a small correction to the error estimates of LP. The typical error per measurement is 19\% and the characteristic depth of the survey is 84 \hmpc.

\item EFAR \citep{ColSagBur01} is a survey of 85 clusters and groups, based on the FP distance indicator. The EFAR survey was not intended to measure the dipole moment, but rather to examine peculiar velocities in two superclusters:  Hercules-Corona Borealis and Perseus-Pisces-Cetus at a distance of $\sim 120$ \hmpc. As a result of this strategy, the coverage is far from all-sky.  The characteristic depth is 93 \hmpc.

\item \cite{Wil99b} is a Tully-Fisher based survey of 15 clusters with a characteristic depth of 111 \hmpc.

\end{itemize}

In addition to treating each of the above surveys independently, it is also interesting to combine them into supersets. The distance range spanned by the surveys is rather heterogeneous, however. Essentially the surveys fall into two categories: dense, relatively shallow surveys of nearby field galaxies or small groups (SBF, ENEAR, SFI++) and sparser but deeper surveys of clusters (EFAR, SC, SMAC, Willick).   The SN sample straddles a range of depths, but is rather sparse, so we associate it with the latter category. However, the large numbers of objects in SFI++ dominate all samples, hence our superset labelled SHALLOW consists of SBF and ENEAR only, and we combine SFI field and group samples separately into a second shallow set labelled SFI++. The DEEP sample includes all other surveys, except for LP (see Section~\ref{sec:consist}). 

Finally, we also combine all surveys (except for LP) into a master catalogue labelled ``COMPOSITE''. The COMPOSITE catalogue has a characteristic depth of 33 \hmpc\ and is based on 4481 peculiar velocity measurements, making it the largest peculiar velocity catalogue studied to date.

\section{Velocity Moments}
\label{sec:mom}

The statistics of individually measured galaxy or cluster peculiar velocities $S_n$ are not described well by linear theory due to the existence of nonlinear flows on small scales.     This problem is typically solved by forming moments as linear combinations of peculiar velocities,  $u_a = \sum_n w_{a,n}S_n$, where $w_{a,n}$ are a set of weights that specify the composition of the $a$th moment.  For a proper choice of weights,  and for a peculiar velocity survey that densely occupies a large volume of space, moments can be formed that are relatively insensitive to small scale motions and thus can be treated by linear theory;  small scale motions are essentially averaged out in the summation.

By far the most common moments used in the analysis of peculiar velocity surveys are the three components of the bulk flow vector.    The bulk flow represents the net motion of the survey volume as a whole as traced by the galaxies occupying it.    For an idealized survey, consisting of positions ${\bf r}_n$ and exact line-of-sight velocities $v_n$ for a set of $N$ galaxies or clusters, the bulk flow vector components $U_i$ are just averages over the projections of the radial velocities onto the three coordinate axis directions, so that the weights for $U_i$ are 
\begin{equation}
w_{i,n} =  {\bf \hat x}_i\cdot{\bf \hat r}_n/N.   
\label{eq:idw}
\end{equation}

For a more realistic survey, the measured line-of-sight velocity is assumed to have the form $S_n = v_n + \delta_n$, where $\delta_n$ is a drawn from a Gaussian distribution with variance $\sigma_n^2 + \sigma_*^2$.   Here $\sigma_n$ is the measurement error of the $n$th galaxy in the survey and $\sigma_*$ is the velocity noise, which accounts for smaller-scale flows not included in our model.    \cite{Kai88} has shown that the weights for the maximum likelihood estimate for the bulk flow components, which we will refer to as the MLE weights,  are
\begin{equation}
w_{i,n} = A_{ij}^{-1} \sum_m{{\bf \hat x}_j\cdot{\bf \hat r}_n\over  \sigma_n^2+\sigma_*^2},
\end{equation}
where
\begin{equation}
A_{ij} = \sum_m {({\bf \hat x}_i\cdot{\bf \hat r}_m)({\bf \hat x}_j\cdot{\bf \hat r}_m)   \over  \sigma_m^2+\sigma_*^2}.
\end{equation}

The statistics for velocity moments can be obtained directly from the formulae for individual velocities obtained from linear theory.  For example, the covariance matrix for a set of moments $u_a$ formed from the velocities $S_n$ of a survey is given by
\begin{eqnarray}
\langle u_au_b\rangle &=& \langle \left(\sum_n w_{a,n}S_n\right)\left(\sum_m w_{b,m}S_m\right)\rangle \nonumber\\
&=& \sum_{n,m} w_{a,n} w_{b,m}\langle S_nS_m\rangle\nonumber\\
&=& \sum_{n,m}w_{a,n} w_{b,m} G_{nm}.
\label{eq:cov}
\end{eqnarray}

The covariance matrix for the individual measured velocities $G_{nm}=\langle S_nS_m\rangle$ can be written in terms of the velocity field ${\bf v}(\bf{r})$ as
\begin{eqnarray}
G_{nm} &=& \langle v_n v_m\rangle
+ \delta_{nm}(\sigma_*^2 + \sigma_n^2)\nonumber\\
&=&\langle {\bf \hat r}_n\cdot {\bf v}({\bf r}_n)\ \   {\bf\hat r}_m\cdot {\bf v}({\bf r}_m)\rangle
+ \delta_{nm}(\sigma_*^2 + \sigma_n^2).
\label{eq:galv}
\end{eqnarray}
In linear theory the first term can be expressed as an integral over the density power spectrum $P(k)$, 
\begin{equation}
\langle {\bf \hat r}_n\cdot {\bf v}({\bf r}_n)\ \   {\bf\hat r}_m\cdot {\bf v}({\bf r}_m)\rangle
=  {\Omega_{m}^{1.1}\over 2\pi^2}\int   dk\  P(k)f_{mn}(k),
\label{eq:vcov}
\end{equation}
where the function $f_{mn}(k)$ is the angle averaged window function 
 \begin{eqnarray}
 f_{mn}(k) = \int &&{d^2{\hat k}\over 4\pi} \left( {\bf \hat r}_n\cdot {\bf \hat k} \right)\left( {\bf \hat r}_m\cdot {\bf \hat k} \right) \nonumber\\
 &&\times  \exp \left(ik{\bf \hat k}\cdot ({\bf r}_n - {\bf r}_m)\right).
\end{eqnarray}

Plugging Eq. (\ref{eq:galv}) into Eq. (\ref{eq:cov}) and using equation (\ref{eq:vcov}),  the covariance matrix of the moments reduces to two terms,
 \begin{equation}
 R_{ab} = R^{(v)}_{ab} +  R^{(\epsilon)}_{ab} .
\label{eq:rab}
 \end{equation}
 The second term, called the ``noise" term, is given by
 \begin{equation}
 R^{(\epsilon)}_{ab}  = \sum_{n} w_{a,n}w_{b,n}\left( \sigma_n^2 + \sigma_*^2\right).
  \end{equation}
 The  first term is given as an integral over the matter fluctuation power spectrum, $P(k)$, 
 \begin{equation} 
 R^{(v)}_{ab}  = {\Omega_{m}^{1.1}\over 2\pi^2}\int_0^\infty
 dk \ \  {\cal W}^2_{ab}(k)P(k),
 \label{eq:covv}
 \end{equation}
 where the angle-averaged tensor window function is 
 \begin{eqnarray}
 {\cal W}^2_{ab} (k)= \sum_{n,m} w_{a,n} w_{b,m}&\int& {d^2{\hat k}\over 4\pi}\ \left({\bf \hat r}_n\cdot {\bf \hat k}\ \ {\bf \hat r}_m\cdot {\bf \hat k}\right)\nonumber\\
 &&\times \exp\left(i{\bf k}\cdot ({\bf r}_n- {\bf r}_m)\right).\nonumber\\
\label{eq:win}
\end{eqnarray}

For the case $a=b$, Eq. (\ref{eq:win}) gives the angle-averaged window function for the moment $u_a$.    This window function tells us which scales contribute to the value of the moment.   For moments that are measures of a component of the bulk flow,  the window function should have the value of $1/3$ at $k=0$.   This ensures that if the flow in the survey volume were uniform, \ie all the power was at $k\sim 0$, the variance of the moment would correctly be $1/3$ of the flow variance.  Ideally, the window function for any useful moment should have a small amplitude at values of $k$ corresponding to nonlinear scales; thus moments that measure the bulk flow components tend to have a peak at $k=0$ with the amplitude falling toward a plateau as $k$ increases.

For the MLE weights, the bulk flow moment window functions are determined by the spatial distribution of objects as well as their associated velocity uncertainties.   Given the fact that most surveys have relatively more objects at smaller distance, and that the measurement uncertainty increases rapidly with distance,  the window functions found using the MLE weights tend to have broader peaks then one might naively expect given the depth of the survey, leading to bulk flow moments that are sensitive to motions on somewhat smaller scales than the diameter of the survey.   

Since our goal here is to study motions on the largest scales possible, we require bulk flow moments whose window functions have as narrow a peak as possible, also being small in amplitude outside the peak.     Given that the moments found using the MLE weights for a typical survey do not generally meet these criteria, we have formulated a new method for calculating weights for moments that essentially allow us to ``design" the moment's window function, subject, of course, to the distribution and uncertainties of the data that is available.    

We begin by considering an idealized survey whose bulk flow components  $U_i$ have the desired window function.   Here we will use an ideal survey which consists of a very large number of objects isotropically  distributed with a Gaussian falloff in density,  $n(r) \propto \exp(-r^2/2R_I^2)$, where $R_I$ specifies the depth of the survey whose velocity is measured exactly.    For this survey, Eq.  (\ref{eq:idw}) gives the weights for the bulk flow components, which will all have the same window function due to isotropy.     
Now, suppose that we have a galaxy or cluster survey consisting of positions ${\bf r}_n$ and measured line-of-sight velocities $S_i$ with associated measurement errors $\sigma_n$.   We would like to find the weights $w_{i,n}$ that specify the three moments $u_i = \sum_n w_{i,n}S_n$ that minimize the average variance, $\langle (u_i-U_i)^2\rangle$.    We will call these the minimum variance, or MV weights.  The MV moments $u_i$ calculated from these weights will then be the best estimate of the bulk flow of the ideal survey, if it were to exist, that can be obtained from the available data.     We also expect that, within limits that will be described more fully below, the window functions of the $u_i$ will match those of the ideal survey.

In order to calculate the weights $w_{i,n}$, we first expand out the variance in terms of the weights,
\begin{eqnarray}
\langle (u_i-U_i)^2\rangle &=& \sum_{n,m} w_{i,n}w_{i,m}\langle S_nS_m\rangle + \langle U_i^2\rangle \nonumber\\
&&- 2\sum_n w_{i,n}\langle  U_iv_n\rangle ,
\label{eq:variance}
\end{eqnarray}
where we have used the fact that the measurement error included in $S_n$ is uncorrelated with the bulk flow $U_i$.   

The next step would be to minimize this expression with respect to $w_{i,n}$; however, as discussed above in order to be a proper measure of the bulk flow, the window function of a moment must go to $1/3$ as $k\rightarrow 0$.     From Eq. (\ref{eq:win}), we see that this requires the constraint 
\begin{equation}
 \lim_{k\to 0} {\cal W}^2_{ii} (k)=   \sum_{n,m} w_{i,n} w_{i,m} P_{nm}=1/3,
\label{eq:con}
\end{equation}
where 
\begin{equation}
P_{nm} = \int {d^2{\hat k}\over 4\pi}\ \left({\bf \hat r}_n\cdot {\bf \hat k}\ \ {\bf \hat r}_m\cdot {\bf \hat k}\right).
\label{eq:pdef}
\end{equation}
We enforce this constraint using the Lagrange multiplier method; thus the quantity to be minimized with respect to $w_{i,n}$ becomes
\begin{eqnarray}
&\sum_{n,m}& w_{i,n}w_{i,m}\langle S_nS_m\rangle + \langle U_i^2\rangle - 2\sum_n w_{i,n}\langle u_i U_i\rangle \nonumber\\
&& + \lambda \left(\sum_{n,m} P_{nm}w_{i,n}w_{i,m} - 1/3\right).
\end{eqnarray}

Taking the derivative with respect to $w_{i,n}$ and setting it equal to zero gives
\begin{equation}
\sum_m \left(\langle S_nS_m\rangle+\lambda P_{nm}\right)w_{i,m}  = \langle S_nU_i\rangle,
\end{equation}
which can be written in matrix form as 
\begin{equation}
 \left({\bf G}+\lambda {\bf P}\right){\bf w_i} = {\bf Q_i},
\end{equation}
where ${\bf G}_{nm} = \langle S_nS_m\rangle$ is the individual velocity covariance matrix, the components of ${\bf Q}_i$ are ${\bf Q}_{i,n} = \langle U_iv_n\rangle$, and ${\bf w_i}$ is the $N$-dimensional vector of weights specifying the $i$th moment.   This is easily solved to give
\begin{equation}
{\bf w_i} =  ({\bf G}+\lambda {\bf P})^{-1}{\bf Q_i}.
\end{equation}
This equation, together with the constraint given in Eq. (\ref{eq:con}), allows us to solve for the weights in terms of the covariance matrix ${\bf G}$, the matrix ${\bf P}$ given in Eq. (\ref{eq:pdef}), and the $n$-dimensional vector ${\bf Q_i}$.   Note that the MV weights can be positive or negative.  If the ideal survey consists of $N^\prime$ exact velocities $v_{n^\prime}$ measured at positions ${\bf r}^\prime_{n^\prime}$, then the elements of ${\bf Q_i}$ can be written as
\begin{equation}
{\bf Q}_{i,n} = \langle U_iv_n\rangle 
= \sum_{n^\prime=1}^{N^\prime}  w^\prime_{i,n^\prime}\langle v_{n^\prime} v_n\rangle ,
\end{equation}
where $w^\prime_{i,n^\prime} = {\bf \hat x}_i\cdot{\bf r}_{n^\prime}^\prime/N^\prime$ as discussed above.  The quantity $\langle v_{n^\prime} v_n\rangle$ can be calculated from Eq. (\ref{eq:vcov}) in terms of the positions ${\bf r}^\prime_{n^\prime}$ and ${\bf r}_n$.    In practice, we calculate 
${\bf Q}_{i,n}$ by constructing a simulated ideal survey where positions ${\bf r}^\prime_{n^\prime}$ are selected at random to be isotropic and to have the density $n(r) \propto \exp(-r^2/2R_I^2)$.   For the purposes of this study we have found that $N^\prime = 10^4$ points are sufficient 
for convergence of all relevant quantities.

Note that the weights depend on the spectrum of matter fluctuations, (see Eq.~(\ref{eq:vcov})).   Here we use the power spectrum model given by  \cite{EisHu98}, which explicitly includes the effect of baryons.   In this parametrization, $P(k)\propto k^{n_s}T^2(k)$, where $n_s$ is the spectral index and the transfer function $T(k)$ depends on the parameters $h$, $\Omega_m$, $\Omega_b$, the baryon density parameter, and $\sigma_8$, the amplitude of matter density perturbations on the scale of $8\ h^{-1}$Mpc. 

\section{Results: The MV weights, window functions and moments}
\label{sec:res}
The minimum variance (MV) weights were calculated for the bulk flow component moments using the method described above for each of our catalogues.  Here we will show results for two different ideal survey radii, a relatively shallow survey, $R_I= 20\ h^{-1}$Mpc, and a deep survey, $R_I= 50\ h^{-1}$Mpc. For calculating the weights, we assume the WMAP5 \cite{DunKomNol08} central parameters $\Omega_m=0.258$, $\Omega_b = 0.0441$, $\sigma_8 = 0.796$, $h = 0.719$, and the spectral index $n_s = 0.963$, together with $\sigma_*=150.0$ km/s.   We note that, for all but the sparsest surveys we consider, the values of the weights are insensitive to the specific power spectrum parameters used.

One qualitative way to gauge how a moment constructed in this way matches its ideal counterpart is to compare the window functions as calculated using Eq.~(\ref{eq:win}).    Generally, the larger and more geometrically similar a survey is to the ideal distribution, the better the match will be.   For small surveys and/or those that have a very different spatial distribution than the ideal distribution the match can be quite poor.   Measurement errors also play a large role in determining how good a match is obtained.   Since the quantity $\langle (u_i-U_i)^2\rangle$ that is being minimized includes the noise term $\langle u_i^2\rangle$, the optimal weighting scheme is a compromise between the need to have the moment's window function match the ideal window function and the need to keep the noise small by giving small weights to objects with large measurement errors.
   
\begin{figure*}
     \includegraphics[width= \textwidth]{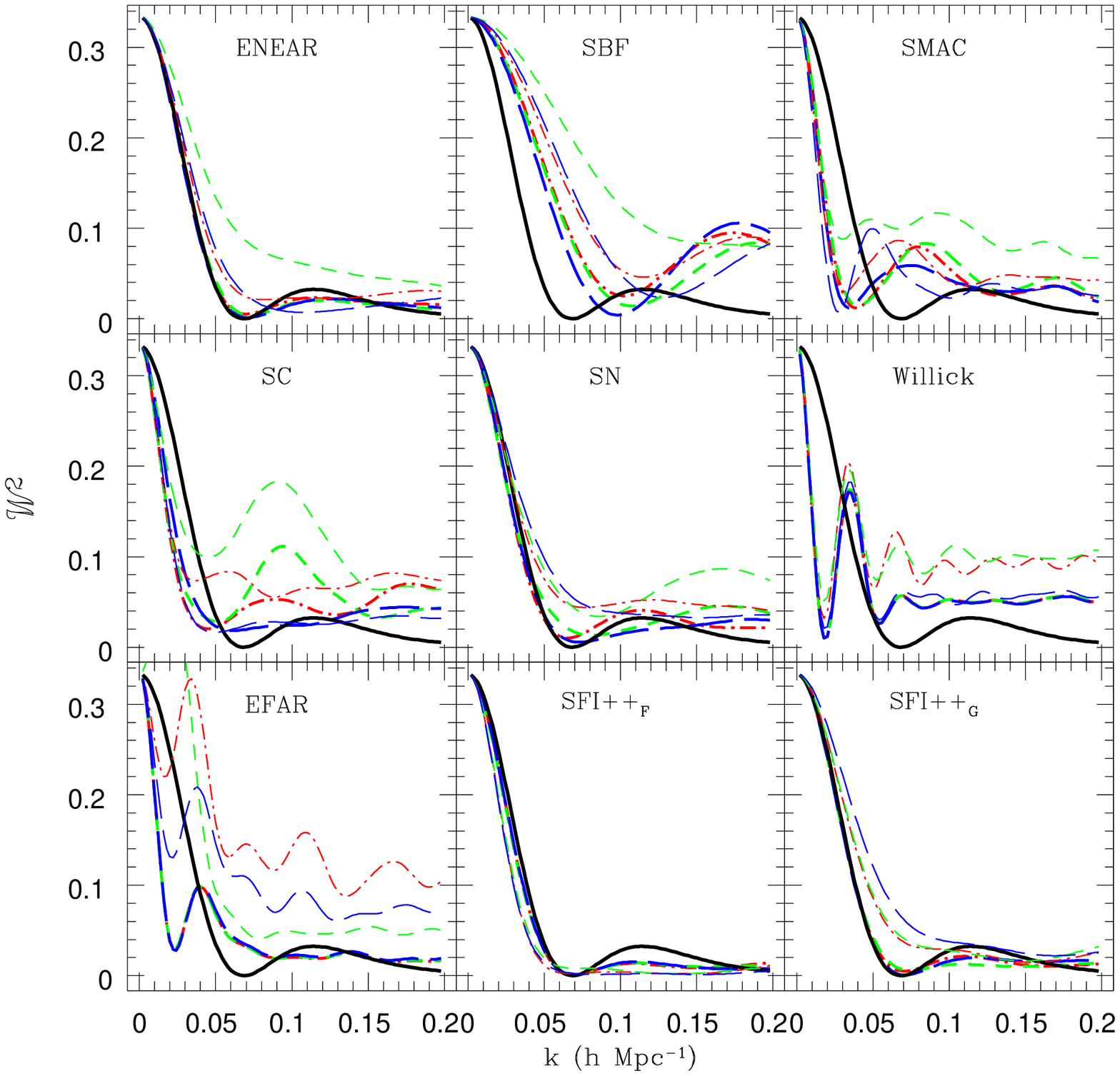}
        \caption{The window functions of the bulk flow component for $R_I=20$\hmpc\  for all of the catalogues we considered. The thick (thin) lines are the window functions for the MV (MLE) bulk flow components. The x,y,z--component are dash-dot, short dash, long dash lines respectively. The thick solid line is the ideal window function (since the ideal survey is isotropic, all component are the same).}
    \label{fig:win20-1}
\end{figure*}

\begin{figure*}
     \includegraphics[width= \textwidth]{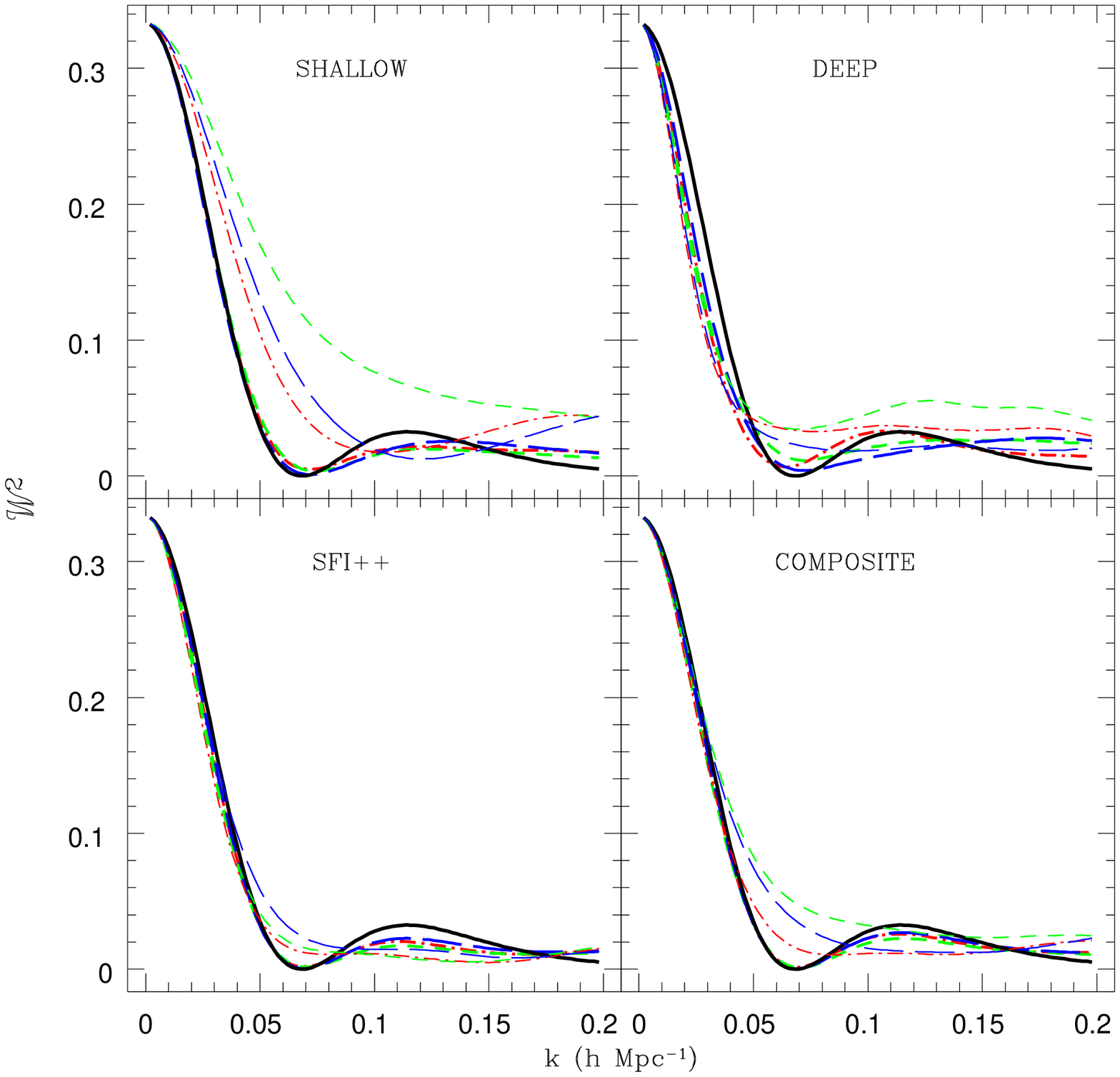}
        \caption{The same as Fig.~\ref{fig:win20-1} for composite surveys}
    \label{fig:win20-2}
\end{figure*}

\begin{figure*}
     \includegraphics[width= \textwidth]{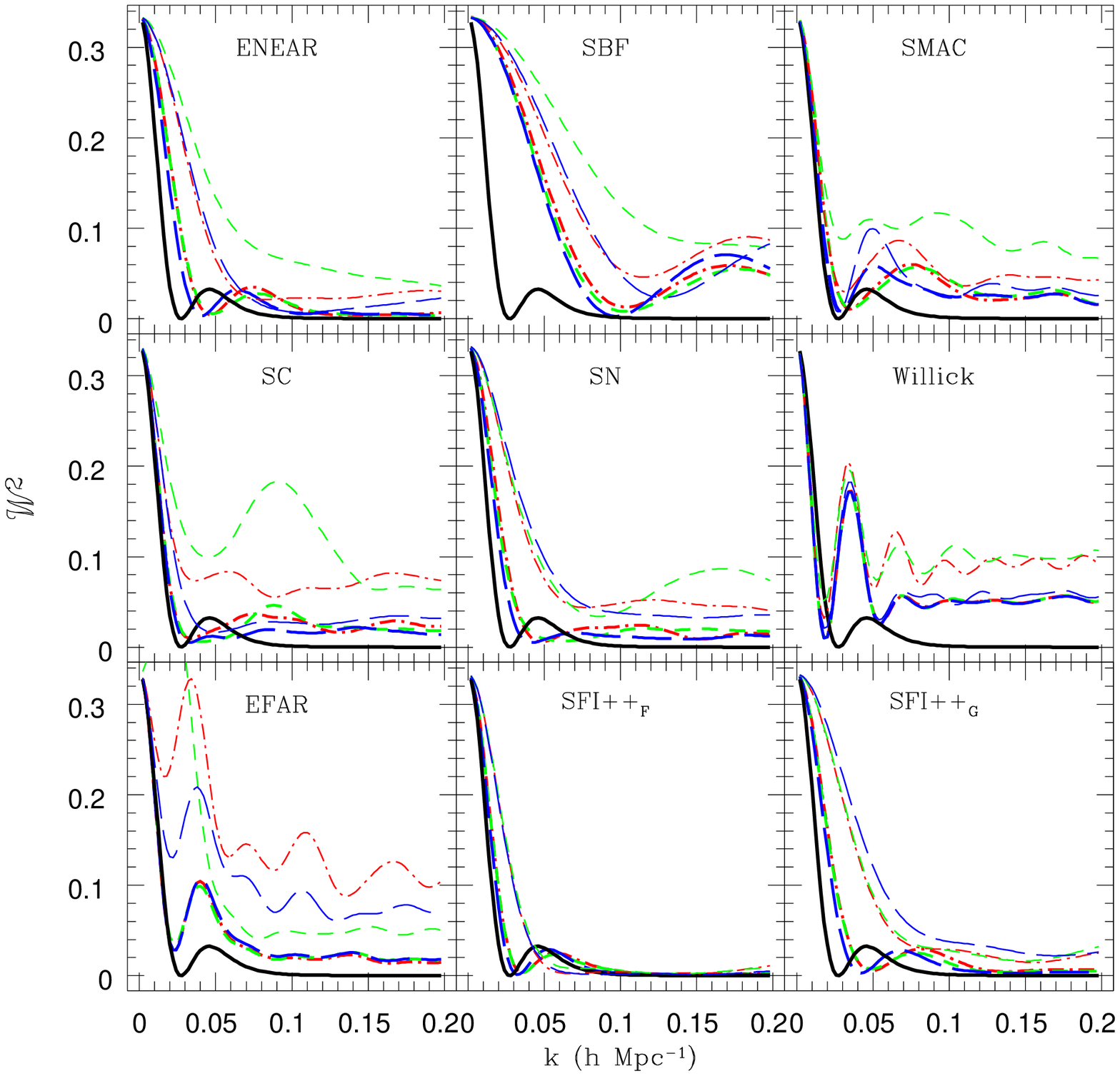}
        \caption{The same as Fig.~\ref{fig:win20-1} for $R_I=50$\hmpc.}
    \label{fig:win50-1}
\end{figure*}

\begin{figure*}
     \includegraphics[width= \textwidth]{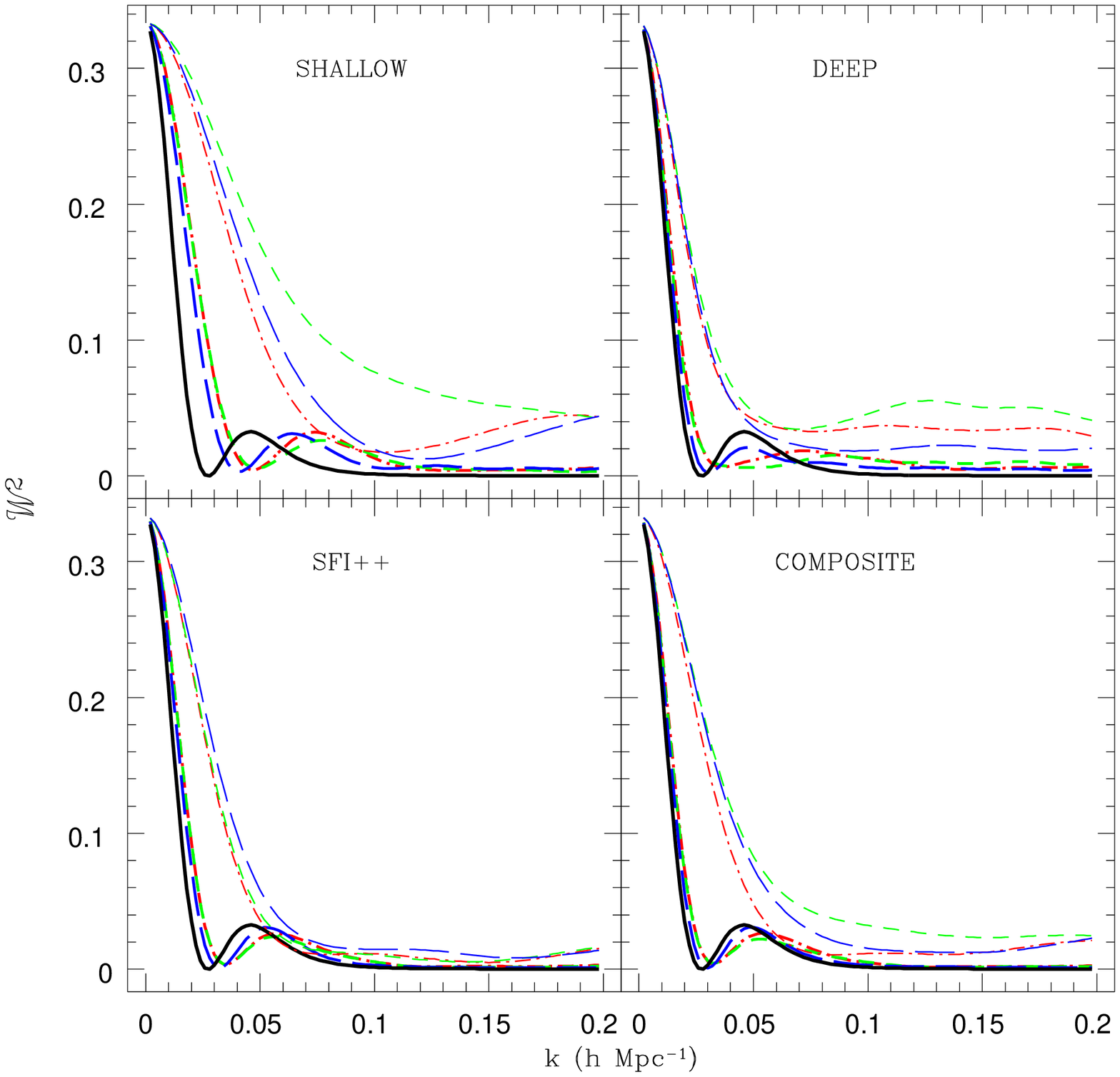}
        \caption{The same as Fig.~\ref{fig:win20-2} for $R_I=50$\hmpc.}
    \label{fig:win50-2}
\end{figure*}

In Figures~\ref{fig:win20-1} -- \ref{fig:win50-2} we show the window functions of the MV bulk flow component moments.   For comparison, we also include the ideal window functions as well as those for the MLE moments for each survey.     As expected, the match between the window functions for the MV moments and the ideal is best for the large surveys and those with small measurement error and similar distribution to the ideal survey.   For the sparse, noisy surveys, the window functions for the MV moments are not very different than those of the MLE moments, differing mostly in the amplitude of the tail of the window function for large $k$.    There are some exceptions: for example, for the $R_I = 50 \hmpc$ case, the SFI++$_G$ and SHALLOW catalogue MV weights approximate a $R \sim20 \hmpc$ Gaussian rather than the desired $R_I = 50 \hmpc$ Gaussian. The COMPOSITE catalogue window function is an excellent match on both 20 and 50 \hmpc\ scales.

There are several ways to quantify how well the moments constructed in this way should agree with their ideal counterparts.   First, from Eq.~(\ref{eq:variance}) we can calculate $\sqrt{ \langle (u_i-U_i)^2\rangle}$.   An alternative measure is to define the correlation coefficient $ {\langle u_i\cdot U_i\rangle/ |u_i||U_i|}$.   A value close to 1 indicates that the moment is an accurate measurement of the bulk flow of the idealized volume.   While both of these measures depend on an assumed power spectrum model, the correlation has the advantage of being dependent only on the shape of the power spectrum and not on its amplitude.   The correlations between the MV moments and the ideal moments given in Table~\ref{tab:bfm} quantify how well they are expected to agree, accounting for both agreement between their window functions and measurement error.    From this table it is clear that while our method works well for surveys which are large or have small errors, the moments it calculates for sparse, deep surveys that have large errors are not strongly correlated with the ideal moments that they are designed to measure.   However, including these surveys in the composite catalogues does improve their correlations, particularly for the case of $R_I=50\ h^{-1}$Mpc.   In particular, the DEEP catalogue does have a strong expected correlation (0.80) with the ideal survey with $R_I=50\ h^{-1}$Mpc, even though each of its component surveys does not. As noted above, the correlation (0.91) of the COMPOSITE survey with the ideal survey $R_I = 50 \hmpc$ is excellent, and the uncertainty due to the mismatch between ideal survey and actual weighted sampling is very small ($\sim 20-30 \kms$).

\begin{figure*}
     \includegraphics[width=\textwidth]{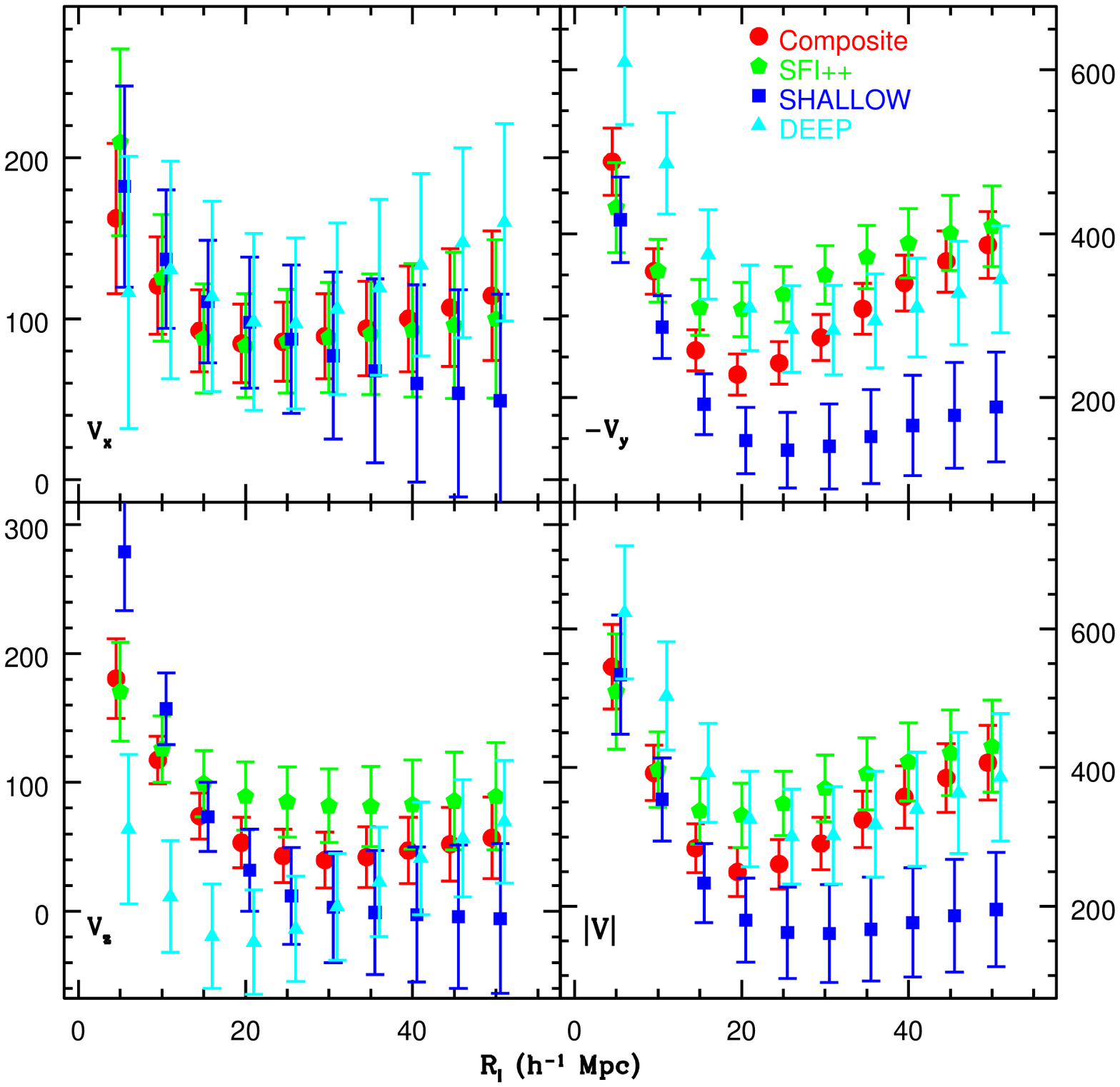}
        \caption{The bulk flow of the composite catalogues as a function of $R_I$. Note that the data points are not independent, but rather are highly correlated. In the Galactic y direction (upper right panel), there is a consistent and robust flow exhibited by all catalogues that probe large scales,  and this is reflected in the BF magnitude (lower right panel).}
    \label{fig:surveybulk}
\end{figure*}

In Table~\ref{tab:bfm} we also show the magnitude of the MV moments, the expected deviation from the ideal moment, $\sqrt{\langle (u_i-U_i)^2\rangle}$, given our power spectrum model.   We also include the measurement noise in parenthesis to show how much of the deviation from the ideal moment is due to measurement noise and how much is due to differences in how the moments probe the window function.    In Table~\ref{tab:bfv} we show the Cartesian components of the MLE moments together with the MV moments for both $R_I=20h^{-1}$Mpc and $R_I=50h^{-1}$Mpc.  The Cartesian components as a function of the Gaussian window $R_I$ for the composite surveys are also shown in Figure \ref{fig:surveybulk}.   The first thing to note from Tables~\ref{tab:bfm} and \ref{tab:bfv} and Figure \ref{fig:surveybulk} is the remarkable consistency between the catalogues as to the value for the bulk flow components.    This consistency will be explored in more detail below.   

\begin{table*}
\caption{Bulk flow amplitudes $|u|$ for the MLE-weighted case and for two choices of $R_I$ for the MV weights.  For the MV moments, the error in parenthesis is the noise error only. The quoted error includes both noise and the difference from the idealized survey geometry.  Also shown are the correlation coefficients $\langle\frac{{\bf u}\cdot {\bf U}}{|{\bf u}||{\bf U}|}\rangle$ between ideal moments and their MLE estimates, where 1 indicates a perfect correlation.}
\begin{tabular}{lccccc}  
\hline \hline
          & MLE weighted bulk flow &  \multicolumn{4}{c}{MV weighted bulk flow}     \\
          &    &  \multicolumn{2}{c}{$R_I=20$\hmpc} & \multicolumn{2}{c}{$R_I=50$\hmpc}     \\
Survey    &  $ |u|$      &$\langle\frac{{\bf u}\cdot {\bf U}}{|{\bf u}||{\bf U}|}\rangle$   &   $|u|$   &$  \langle \frac{{\bf u}\cdot {\bf U}}{|{\bf u}||{\bf U}|}\rangle$   &   $|u|$ \\
          &  km/s        &  & km/s & & km/s \\ \hline \hline
         SBF &   382 $\pm$   82&    0.859 &   360 $\pm$   177 (  103)  &    0.377 &   419 $\pm$   316 (  111)  \\ \hline
           ENEAR &   336 $\pm$   55&    0.946 &   207 $\pm$    89 (   73)  &    0.735 &   205 $\pm$   169 (  112)  \\ \hline
         SHALLOW &   340 $\pm$   48&    0.953 &   180 $\pm$    83 (   66)  &    0.735 &   195 $\pm$   169 (  111)  \\ \hline
\multicolumn{6}{c}{ }  \\ 
              LP &   833 $\pm$  408&    0.188 &  1287 $\pm$   438 (  353)  &    0.259 &  1299 $\pm$   392 (  355)  \\ \hline
 \multicolumn{6}{c}{ }  \\ 
        Willick &  1019 $\pm$  542&    0.013 &    41 $\pm$   609 (  505)  &    0.125 &    76 $\pm$   557 (  505)  \\ \hline
            EFAR &   736 $\pm$  441&    0.099 &   361 $\pm$   410 (  269)  &    0.264 &   375 $\pm$   337 (  268)  \\ \hline
            SMAC &   657 $\pm$  189&    0.447 &   569 $\pm$   290 (  172)  &    0.535 &   578 $\pm$   234 (  179)  \\ \hline
              SC &   132 $\pm$  138&    0.694 &   182 $\pm$   217 (  154)  &    0.579 &   151 $\pm$   204 (  153)  \\ \hline
              SN &   471 $\pm$   76&    0.878 &   426 $\pm$   136 (  106)  &    0.664 &   477 $\pm$   189 (  137)  \\ \hline
            DEEP &   338 $\pm$   60&    0.906 &   326 $\pm$   115 (   86)  &    0.796 &   386 $\pm$   126 (  101)  \\ \hline
\multicolumn{6}{c}{ }  \\ 
       SFI++$_G$ &   370 $\pm$   58&    0.940 &   357 $\pm$    94 (   76)  &    0.719 &   471 $\pm$   172 (  116)  \\ \hline
      SFI++$_F$ &   357 $\pm$   54&    0.949 &   322 $\pm$    85 (   71)  &    0.855 &   375 $\pm$   109 (   88)  \\ \hline
           SFI++ &   364 $\pm$   40&    0.966 &   331 $\pm$    69 (   53)  &    0.870 &   431 $\pm$   102 (   81)  \\ \hline
\multicolumn{6}{c}{ }  \\
       COMPOSITE &   341 $\pm$   27&    0.977 &   249 $\pm$    57 (   40)  &    0.908 &   407 $\pm$    81 (   65)  \\ \hline
\label{tab:bfm}
\end{tabular}
\end{table*} 

\begin{table*}
\caption{Bulk flow vectors for the surveys (in Galactic Cartesian coordinates), for MLE- and MV weights . For the MV moments, the error in parenthesis is the noise error only. The quoted error includes both noise and the difference from the idealized survey geometry. Note that the MV $R_I = 20 \hmpc$ and $R_I = 50 \hmpc$ results for a given survey are not independent.} 
\begin{tabular}{lccccccc}
 \hline \hline
Survey      &  \multicolumn{3}{c}{MLE} &&\multicolumn{3}{c}{MV}\\
& $v_x$ & $v_y$ & $v_z$ &  $R_I$ & $v_x$ & $v_y$ & $v_z$  \\
& (km/sec) & (km/sec) & (km/sec) &  ($h^{-1}$Mpc) & (km/sec) & (km/sec) &  (km/sec) \\ \hline \hline
         SBF &  248 $\pm$   56 & -212 $\pm$   50 &  198 $\pm$   35  &   20     &  209 $\pm$  107 (  64) & -242 $\pm$  110 (  63) &  165 $\pm$   88 (  51)   \\ \cline{5-8}
                        &   &   &  &   50             &  244 $\pm$  188 (  65) & -277 $\pm$  183 (  68) &  198 $\pm$  177 (  59)   \\ \hline
           ENEAR &  179 $\pm$   41 & -281 $\pm$   35 &   41 $\pm$   31  &   20     &  118 $\pm$   51 (  46) & -170 $\pm$   59 (  44) &   -0 $\pm$   42 (  36)   \\ \cline{5-8}
                        &   &   &  &   50             &   56 $\pm$  102 (  67) & -196 $\pm$  105 (  68) &  -20 $\pm$   84 (  59)   \\ \hline
         SHALLOW &  190 $\pm$   33 & -259 $\pm$   29 &  110 $\pm$   23  &   20     &   98 $\pm$   47 (  41) & -148 $\pm$   57 (  41) &   32 $\pm$   39 (  32)   \\ \cline{5-8}
                        &   &   &  &   50             &   49 $\pm$  102 (  66) & -189 $\pm$  105 (  67) &   -6 $\pm$   84 (  58)   \\ \hline
\multicolumn{8}{c} {} \\
              LP &  512 $\pm$  295 &  -99 $\pm$  331 &  649 $\pm$  241  &   20     &  757 $\pm$  257 ( 205) & -735 $\pm$  252 ( 203) &  738 $\pm$  251 ( 204)   \\ \cline{5-8}
                        &   &   &  &   50             &  777 $\pm$  232 ( 206) & -738 $\pm$  229 ( 203) &  735 $\pm$  217 ( 205)   \\ \hline
\multicolumn{8}{c} {} \\
         Willick &   97 $\pm$  392 & -904 $\pm$  406 &  461 $\pm$  321  &   20     &   23 $\pm$  353 ( 292) &   19 $\pm$  352 ( 292) &   29 $\pm$  351 ( 291)   \\ \cline{5-8}
                        &   &   &  &   50             &   43 $\pm$  332 ( 293) &  -34 $\pm$  321 ( 291) &   54 $\pm$  311 ( 291)   \\ \hline
            EFAR &  518 $\pm$  344 &  519 $\pm$  258 &   57 $\pm$  223  &   20     &  250 $\pm$  237 ( 156) & -177 $\pm$  243 ( 155) &  191 $\pm$  231 ( 154)   \\ \cline{5-8}
                        &   &   &  &   50             &  273 $\pm$  195 ( 155) & -148 $\pm$  211 ( 157) &  211 $\pm$  176 ( 153)   \\ \hline
            SMAC & -116 $\pm$  157 & -647 $\pm$  161 &  -20 $\pm$  118  &   20     &  323 $\pm$  169 ( 103) & -414 $\pm$  165 (  95) &  217 $\pm$  169 ( 100)   \\ \cline{5-8}
                        &   &   &  &   50             &  285 $\pm$  140 ( 106) & -460 $\pm$  143 ( 100) &  202 $\pm$  121 ( 104)   \\ \hline
              SC &  120 $\pm$  102 &  -35 $\pm$   91 &   42 $\pm$   68  &   20     &  105 $\pm$  140 ( 105) & -132 $\pm$  134 (  87) &   67 $\pm$   98 (  72)   \\ \cline{5-8}
                        &   &   &  &   50             &  106 $\pm$  125 ( 101) &  -93 $\pm$  131 (  85) &   53 $\pm$   92 (  76)   \\ \hline
              SN &   77 $\pm$   62 & -462 $\pm$   63 &  -52 $\pm$   42  &   20     &   85 $\pm$   81 (  66) & -407 $\pm$   90 (  67) &  -88 $\pm$   62 (  47)   \\ \cline{5-8}
                        &   &   &  &   50             &  160 $\pm$  113 (  80) & -449 $\pm$  121 (  86) &   16 $\pm$   92 (  70)   \\ \hline
            DEEP &  105 $\pm$   49 & -322 $\pm$   48 &   -0 $\pm$   33  &   20     &   98 $\pm$   69 (  55) & -310 $\pm$   75 (  52) &  -24 $\pm$   54 (  41)   \\ \cline{5-8}
                        &   &   &  &   50             &  160 $\pm$   75 (  61) & -345 $\pm$   84 (  65) &   69 $\pm$   56 (  48)   \\ \hline
\multicolumn{8}{c} {} \\
       SFI++$_G$ &  109 $\pm$   41 & -336 $\pm$   42 &  109 $\pm$   27  &   20     &   80 $\pm$   53 (  46) & -334 $\pm$   66 (  49) &  101 $\pm$   42 (  36)   \\ \cline{5-8}
                        &   &   &  &   50             &  109 $\pm$  103 (  69) & -448 $\pm$  107 (  70) &   96 $\pm$   87 (  61)   \\ \hline
      SFI++$_F$ &  123 $\pm$   38 & -327 $\pm$   38 &   78 $\pm$   31  &   20     &   91 $\pm$   46 (  42) & -297 $\pm$   56 (  42) &   84 $\pm$   44 (  38)   \\ \cline{5-8}
                        &   &   &  &   50             &   80 $\pm$   65 (  54) & -351 $\pm$   68 (  53) &  106 $\pm$   54 (  46)   \\ \hline
           SFI++ &  117 $\pm$   28 & -331 $\pm$   28 &   96 $\pm$   20  &   20     &   83 $\pm$   36 (  32) & -308 $\pm$   49 (  33) &   89 $\pm$   32 (  26)   \\ \cline{5-8}
                        &   &   &  &   50             &  100 $\pm$   61 (  49) & -409 $\pm$   65 (  49) &   89 $\pm$   50 (  42)   \\ \hline
\multicolumn{8}{c} {} \\
       COMPOSITE &  137 $\pm$   19 & -301 $\pm$   18 &   83 $\pm$   14  &   20     &   85 $\pm$   27 (  24) & -228 $\pm$   43 (  25) &   53 $\pm$   26 (  19)   \\ \cline{5-8}
                        &   &   &  &   50             &  114 $\pm$   49 (  40) & -387 $\pm$   53 (  41) &   57 $\pm$   37 (  32)   \\ \hline
\label{tab:bfv}
\end{tabular}
\end{table*}

It is important to note that for a given catalogue, the bulk flow moments calculated for different values of $R_I$ do not have independent errors, and thus cannot be strictly compared.   However, surprisingly, nearly all of the surveys we studied have a larger amplitude flow for $R_I=50\ h^{-1}$Mpc than for $R_I=20\ h^{-1}$Mpc (see Tables~\ref{tab:bfm} and \ref{tab:bfv} and Figure~\ref{fig:surveybulk}).   It is particularly compelling that this is the case for both the SFI++ and the DEEP catalogues,  which are completely independent and both of which have relatively small measurement errors on their bulk flow moments.   A notable exception to this trend is the SC survey. This leads to the question of whether different surveys are mutually compatible, a topic addressed in the following section.

\section{Consistency of survey bulk flows}
\label{sec:consist}

\subsection{Method}

We now consider the consistency of the bulk flows measured by different surveys.    While this has been done previously using the MLE moments, our MV moments have advantages.   In particular, while MLE bulk flow moments depend on the details of a survey and are not necessarily comparable between different surveys,  the MV moments have been designed to approximate the same window function regardless of the survey particulars.  This means that the theoretical difference of the two moments will be as small as possible, making the comparison more rigorous.   

Following \citet{WatFel95}, we quantify the agreement between two surveys, say survey $A$ and $B$, by calculating the covariance matrix for the difference in value of the bulk flow moments $u_i^A$ and $u_i^B$ of the two surveys,
\begin{eqnarray}
R^{A-B}_{ij}&=&  \langle (u_i^A - u_i^B)(u_j^A - u_j^B)\rangle\nonumber\\
&=& R^A_{ij} + R^B_{ij}-R^{AB}_{ij}-R^{AB}_{ji},
 \end{eqnarray}
where the cross-terms are given by
\begin{equation} 
 R^{AB}_{ij}  = {\Omega_{m}^{1.1}\over 2\pi^2}\int_0^\infty
 dk \ \ ({\cal W}^{AB})^2_{ij}(k)P(k),
\label{eq:covv-diff}
\end{equation}
and
 \begin{eqnarray}
({\cal W}^{AB})^2_{ij} (k)&=&
  \sum_{n,m} w_{i,n}^A\  w_{j,m}^B \int {d^2{\hat k}\over 4\pi}\ \left({\bf \hat r}^A_n\cdot {\bf \hat k}\ \ {\bf \hat r}^B_m\cdot {\bf \hat k}\right)\nonumber\\
&& \times\exp\left(i{\bf k}\cdot ({\bf r}^A_n- {\bf r}^B_m)\right). 
\end{eqnarray}
Agreement of the moments from $A$ and $B$ can then be quantified by a $\chi^2$ for 3 degrees of freedom given by
\begin{equation}
\chi^2 = \sum_{i,j} (u^A_i-u^B_i)(R^{A-B}_{ij})^{-1}(u^A_j-u^B_j).
\label{eq:chisq}
\end{equation}

\subsection{Results}
In order to quantify the agreement between the catalogues, we calculated the $\chi^2$ for three degrees of freedom, as defined in Eq.~\ref{eq:chisq}, for the difference in bulk flow of a given catalogue and that of the composite catalogue \emph{with the given catalogue removed}. Note also that since the covariance is model dependent, the results are given for the WMAP5 central parameters.  The results are given in Table~\ref{tab:consist}.

\begin{table*}
\caption{$\chi^2$ for 3 DoF for composite-surveys for $\Omega_m=0.258$. If the $\chi^2$ value is greater than 7.8, the two surveys disagree at the 95{\%} confidence level.} 
\begin{tabular}{lccccc}
\hline \hline
&   $R_I=$ & \multicolumn{2}{c}{20\hmpc}& \multicolumn{2}{c}{50\hmpc} \\
Survey &  $\sigma_8=$ &   0.796  &  1.0 &   0.796  &  1.0 \\  \hline \hline 
SBF-ENEAR & &  4.036  &  3.026  & 2.828  &  2.008  \\ \hline
\multicolumn{6}{c}{ }  \\
DEEP-SMAC & &  3.288  &  2.558  & 2.493  &  2.135  \\ \hline
DEEP-SC & &  6.351  &  5.071  & 5.980  &  4.817  \\ \hline
DEEP-SN & &  4.963  &  3.929  & 2.615  &  2.137  \\ \hline
DEEP-Willick & &  2.776  &  1.831  & 4.242  &  2.963  \\ \hline
DEEP-EFAR & &  3.230  &  2.182  & 8.387  &  6.241  \\ \hline
\multicolumn{6}{c}{ }  \\
SFI++$_F$-SFI++$_G$ & &  0.430  &  0.401  & 0.981  &  0.843  \\ \hline
\multicolumn{6}{c}{ }  \\
DEEP-SHALLOW & &  4.466  &  3.774  & 3.595  &  2.993  \\ \hline
DEEP-SFI++ & &  3.864  &  3.278  & 1.030  &  0.907  \\ \hline
SFI++-SHALLOW & & 11.300  & 10.446  & 6.090  &  5.000  \\ \hline
\multicolumn{6}{c}{ }  \\
COMPOSITE-ENEAR & &  6.754  &  6.339  & 5.714  &  4.490  \\ \hline
COMPOSITE-SBF & &  2.929  &  2.144  & 2.474  &  1.682  \\ \hline
COMPOSITE-SHALLOW & &  7.400  &  6.843  & 5.999  &  4.712  \\ \hline
\multicolumn{6}{c}{ }  \\
COMPOSITE-SMAC & &  2.739  &  2.120  & 1.939  &  1.607  \\ \hline
COMPOSITE-SC & &  0.605  &  0.447  & 5.293  &  3.922  \\ \hline
COMPOSITE-SN & &  8.104  &  6.664  & 0.975  &  0.808  \\ \hline
COMPOSITE-Willick & &  2.012  &  1.289  & 6.399  &  4.342  \\ \hline
COMPOSITE-EFAR & &  1.235  &  0.827  & 9.604  &  6.782  \\ \hline
COMPOSITE-DEEP & &  2.789  &  2.270  & 1.061  &  0.964  \\ \hline
\multicolumn{6}{c}{ }  \\
COMPOSITE-SFI++$_F$ & &  4.558  &  4.297  & 0.553  &  0.514  \\ \hline
COMPOSITE-SFI++$_G$ & &  6.982  &  6.443  & 0.929  &  0.749  \\ \hline
COMPOSITE-SFI++ & & 12.141  & 11.283  & 2.641  &  2.412  \\ \hline
\multicolumn{6}{c}{ }  \\
SHALLOW-LP & &  8.756  &  7.833  & 10.796  &  9.847  \\ \hline
DEEP-LP & &  9.818  &  8.351  & 10.135  &  8.964  \\ \hline
SFI++-LP & &  8.025  &  6.885  & 11.541  &  9.714  \\ \hline
COMPOSITE-LP & &  8.405  &  7.388  & 12.344  & 10.412  \\ \hline
\hline
\label{tab:consist}
\end{tabular} 
\end{table*}

From Table~\ref{tab:consist}, we see that the MV-weighted bulk flow of LP disagrees with that of the COMPOSITE catalogue on both 20\hmpc\ and 50\hmpc\ scales. The level of disagreement on the latter scale, for $\sigma_8= 0.796$ corresponds to 99\% CL, although this would drop to 98\% if $\sigma_8$ were as high as 1.   This is in agreement with previous analyses based on MLE-weighted moments \citep{Hud03}. There are independent reasons to question the LP results: \cite{HudSmiLuc04} compared, cluster by cluster, the distance to the brightest cluster galaxy derived by LP to that derived from the FP for other cluster members, and found that in a few cases, these distances differed significantly, in the sense that the LP BCG distance was too large (i.e. the BCG was fainter than expected). They found that all of the discrepant BCGs for which HST images were available showed strong evidence for dust. For these reasons, we have chosen to reject LP from the Composite catalogues.

On the 50 \hmpc\ scale, the next most discrepant dataset is EFAR, which disagrees with the COMPOSITE catalogue at the 98\% level, if $\sigma_8= 0.796$. The disagreement is more model-dependent, however, than is the case with LP. If, for example, $\sigma_8$ were as high as 1, the disagreement would drop to only 92\% C.L. Thus we choose to retain EFAR, as well as all of the other catalogues except LP, as part of the COMPOSITE catalogue. On the 20 \hmpc\ scale, we note there is some tension between the SHALLOW catalogue (dominated by ENEAR) and the SFI++ catalogue.  As noted above, on the largest scales, the DEEP catalogue and the SFI++ catalogue are in excellent agreement and both independently show a significant, large-scale flow.

\subsection{Comparison with kinetic Sunyaev-Zel'dovich bulk flow}
\label{sec:ksz}

Recently\footnote{After this paper was submitted},
\cite{KasAtrKoc08b,KasAtrKoc08a}, have claimed a detection of the bulk
flow from the dipole of the CMB observed behind clusters of galaxies,
with amplitude ($2.8 \pm 0.7 \mu$K) and direction towards
\lberr{283}{14}{12}{14}.   They interpret this as a dipole in the kinetic
Sunyaev-Zel'dovich effect. The conversion from $\mu$K to km/s has some systematic uncertainty, but the authors interpret the bulk flow to be between $600 \kms$ and $1000 \kms$. 

Our bulk flow result is in excellent directional agreement
(6$^{\circ}$) with that found by \cite{KasAtrKoc08a}. The amplitude of
their flow ($1000 \kms$) is considerably higher, but would be compatible if systematic and random errors reduced the Kashlinksy et al.\ result to $\sim 400-500 \kms$.  However, we note that their sample is very much deeper than ours. Whereas our signal arises from within a volume of radius $\sim 100 \hmpc$ ($z < 0.03$), their signal is detected on much larger physical scales, with most of the signal arising from the shell in the range $0.04 < z < 0.2$ ($120 \hmpc < r < 600 \hmpc$) .

\section{Implications for Cosmography}
\label{sec:implic}

Our robust measurement of a bulk flow of $407\pm81$ km/s toward \lb{287\arcdeg\pm9}{8\arcdeg\pm6} for a Gaussian window with $R_I = 50 \hmpc$ suggests that roughly 50\% of the Local Group's motion is generated by structures beyond this depth.  This is in good agreement with the value of $366\pm125$ \kms\ toward \lb{300}{13} within a 60 \hmpc\ top-hat sphere, found by \cite{HofEldZar01} based on the Mark III peculiar velocity catalogue.

One way to locate the physical sources of peculiar velocities is through all-sky maps of densities of galaxies, with the assumption that one can use this map as a proxy for that of the dark matter mass, allowing for a bias factor, $b$, between galaxies and dark matter. The degenerate combination of prefactors that scale Eq. \ref{eq:peculiar} is then $\beta = f(\Om)/b$. Having mapped the density field, one can then predict the peculiar motion of the LG and other galaxies.  By comparing these predictions with observed peculiar velocities, one can solve for both $\beta$ and for the bulk motion arising from sources at depths larger than the galaxy survey.  For example, \cite{Hud94b} mapped the density field of optically-selected galaxies within 80 \hmpc, a top-hat radius that is a close match to our 50 \hmpc\ Gaussian. By comparing the predicted peculiar velocities to the old Mark II peculiar velocity data set, he found $\beta = 0.5\pm0.06$ and, more importantly, that the residual motion, arising from sources outside 80 \hmpc, was $405\pm45$ \kms\ toward \lb{292}{7}. This residual motion is remarkably consistent with the result found in this paper in scale, amplitude and direction (within $5\deg$), although it was derived from a completely different data set with different methodology and assumptions. Similarly, \cite{PikHud05}, using a different galaxy density field based on 2MASS photometry and published redshifts, found a slightly lower residual motion of $271\pm104$ \kms toward \lb{300}{15}. This result is also consistent with the $R_I = 50\hmpc$ bulk flow found here.  Redshift surveys of IRAS-selected galaxies \citep{StrYahDav92, RowShaOli00} give very similar results for the residual flow beyond 60 \hmpc, provided $\beta_{IRAS}$ is $\sim 0.5$, as found in direct density-velocity comparisons \citep[see][for a summary table]{PikHud05}.  These independent checks suggest that the large bulk flow motion is consistent with the \emph{absence of} sufficiently massive attractors in the nearby ($r < 80$ \hmpc) Universe.

Neither of the above mentioned galaxy surveys is deep enough to detect the physical source(s) of the 407 \kms\ motion, which must lie beyond $\sim 50 \hmpc$.  The only all-sky \emph{galaxy} survey that reaches much greater depths is the IRAS PSCz survey \citep{BraTeoFre99}. This shows little evidence of important attractors between 60 \hmpc\ and 180 \hmpc, with the exception of the Shapley Concentration, which is relatively weak in IRAS. This issue was studied in more detail by \cite{HudSmiLuc04}, who argued that, to explain the SMAC survey bulk motion, sources generating $\sim 200$ \kms motion must be added to the IRAS-PSCz density field. They attributed this to a combination of (i) sources in the Galactic Plane, (ii) sources beyond the 200 \hmpc\ depth of the PSCz, and (iii) to the undercounting of densest regions of the Shapley Concentration and the Horologium-Reticulum supercluster. The interested reader is referred to Sec 5.4 of \cite{HudSmiLuc04} for further details.

On the other hand, if we assume that clusters of galaxies trace the large-scale density field (perhaps with a higher biasing factor) then one can use an all-sky survey of clusters \citep{KocEbeMul04} as a predictor of the velocity field. This survey suggests that, if clusters trace the mass, the Shapley Concentration and related very large-scale structures may play an important role, with as much as $\sim 300 \kms$ arising from large scales, in approximate agreement with the large flow found here.

Finally, it is worth noting that the large value of the residual motion implies that there are significant velocities generated by very-large scale structures and that this in turn has implications for the impact of such structures as a ``foreground'' for calibrating SNe \citep{CooCal06, HuiGre06, NeiHudCon07, AbaLah08} and for its effects on the CMB such as the Integrated Sachs-Wolfe Effect \citep{FosDor07}.

\section{Comparison with Cosmological Models}
\label{sec:compare}

As discussed above, the large-scale flow is directly sensitive to the large scales of the matter power spectrum, and so one can compare the observed value of the flow to that expected for a given cosmological model. Assuming that the Local Group does not occupy a special location in the Universe,  one can calculate the expected mean flow and variance of bulk flow measurements taken at different locations.  The former quantity is zero and the latter is the quantity of interest: it depends on the weights and sample geometry and is given by the covariance matrix $R_{ij}$ (Eq. \ref{eq:rab}), which depends on the measurement noise and the power spectrum.  This allows us to compare, in a frequentist sense, the observed bulk flow moments with the cosmological expectation. In particular, we can calculate the $\chi^2$ for three degrees of freedom corresponding to the 3 moments, as given by
\begin{equation}
\chi^2 = \sum_{i,j} u_i R_{ij}^{-1}u_j,
\label{eq:chi}
\end{equation}
where $i$ and $j$ both go from 1 to 3 to specify the bulk flow components and the covariance matrix $R_{ij}$ is calculated as described above for a specific set of values for the cosmological parameters.   Here we use the $\Lambda$CDM power spectrum model of \cite{EisHu98} 
and the WMAP5  central parameters as described above.
While this statistic has been calculated previously for MLE bulk flow moments, the advantage of the new MV moments is that, for the case of $R_I= 50h^{-1}$Mpc, they have been designed to be sensitive {\it only} to scales of order 100$h^{-1}$Mpc and larger.    Thus we will be able to probe these scales without having to worry about the influence of smaller scales.    Further, by isolating the very large scale motions, we will see that we will be able to put stronger constraints on  power spectrum parameters.

\begin{table*}
\caption {Expected bulk flows and observed $\chi2$ for 3 bulk flow moments ($R_I = 50h^{-1}$Mpc), as a function of cosmological parameters} 
\begin{tabular}{lcccccc}
\hline \hline
& \multicolumn{2}{c}{ML} &   \multicolumn{2}{c}{BC} \\
& $\Omega_m=0.258$ &$\sigma_8=0.796 $ & 
$\Omega_m=0.262$ &$\sigma_8=0.863 $ \\ 
Expected 1D r.m.s.\ & \multicolumn{2}{c}{112 km/s}   &   \multicolumn{2}{c}{121 km/s}\\ \hline \hline
Survey & $\chi2$ &$P(>\chi2) $ & $\chi2$ & $P(>\chi2)$  \\ \hline
SHALLOW        &  1.54 & 0.6731     &  1.37 & 0.7126  \\
DEEP                &  7.54 & 0.0565      &  6.76 & 0.0800 \\
SFI++                & 11.23 & 0.0105     &  9.92 & 0.0193  \\
COMPOSITE   & 11.52 & 0.0092     & 10.15 & 0.0173 \\
\hline
\label{tab:chisq}
\end{tabular}
\end{table*} 

In Table~\ref{tab:chisq} we show the expected r.m.s. bulk flow for the WMAP5 ($\Omega_m,\sigma_8$) parameters for the COMPOSITE catalogue at a scale of $R_I = 50 \hmpc$. As can be seen by comparing Table~\ref{tab:chisq} with the values in Table~\ref{tab:bfm}, the measured and expected values differ significantly. Quantifying the disagreement, we find, for the WMAP5 central parameters, find that $\chi^2 = 11.52$. The probability of observing a bulk of flow this high an amplitude, in a Universe described by a WMAP5-normalized $\Lambda$CDM model, is only 0.9\%. 

\begin{figure}
\includegraphics[width= \columnwidth]{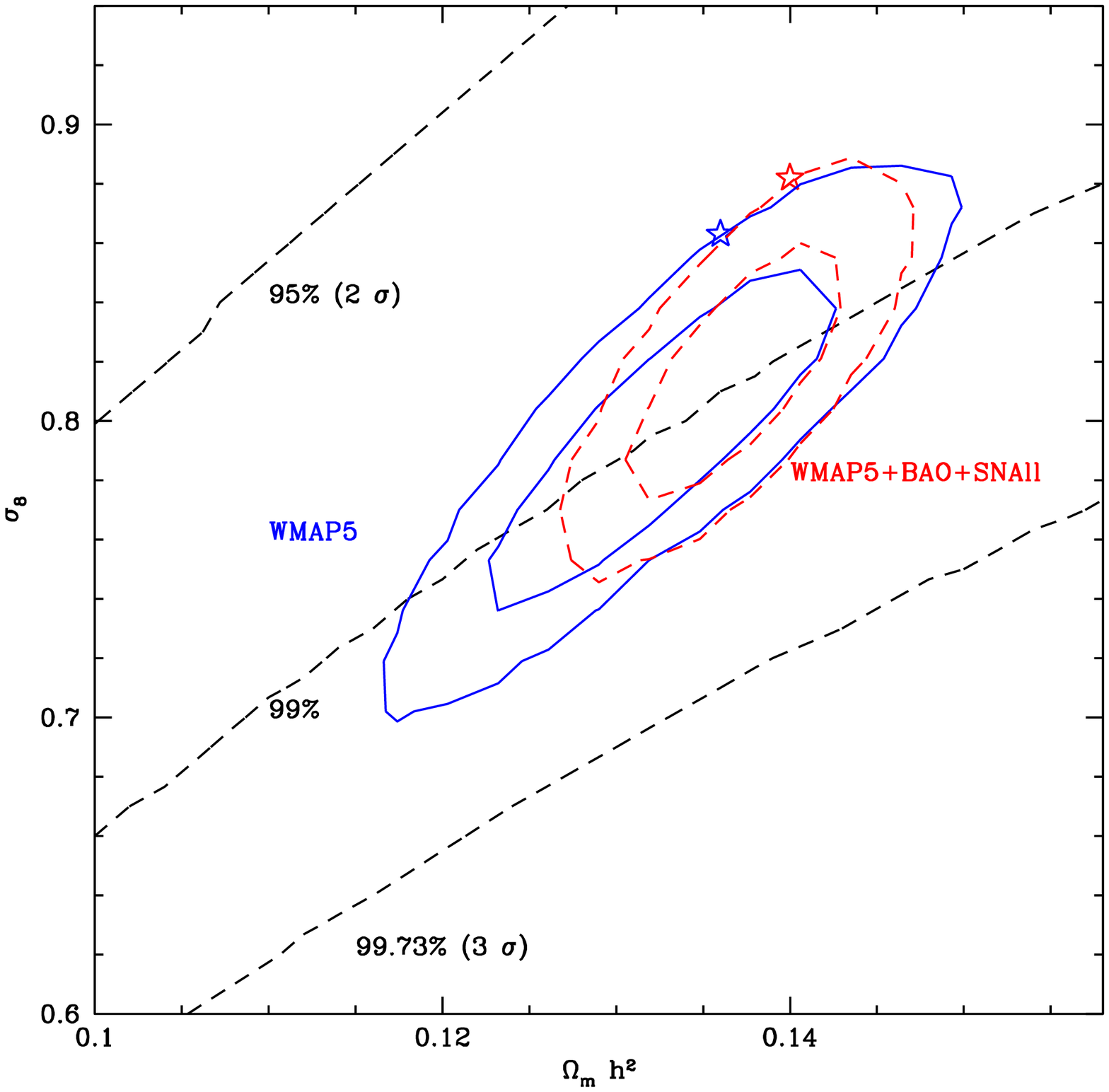}
\caption{
The $\chi^2$-based confidence levels for the MV COMPOSITE survey ($R_I = 50 \hmpc$), given the observed flow, are shown by the black dashed lines at the 95\%, 99\% and 99.73\% levels from top to bottom. Also shown are the WMAP 68\% and 95\% confidence limits from \protect\cite{DunKomNol08}  (blue solid)  well as for WMAP5+BAO+SN \protect\citep{KomDunNol08}  (red dashed). The stars indicate the regions of the WMAP5 parameter space that maximize the bulk flow variance, and hence minimize the $\chi^2$.}
\label{fig:omegamhhsig}
\end{figure}

In order to assess the effect of  the uncertainties in the WMAP5 parameters, we have explored further the multi-dimensional cosmological  parameter space. Our covariance matrix $R_{ij}$ is dominated by the cosmic variance term (typically $\sim 100 \kms$) and not by the noise term  (typically $\sim 40 \kms$) which has a small effect when added in quadrature.  Thus, to a good approximation, $R_{ij}$ should scale with the amplitude of the power spectrum, parametrized by $\sigma_8$. The cosmic variance term also depends on the $f(\Om)$ prefactor and the power spectrum shape parameter, $\Gamma$, which on the large scales probed here is well approximated by $\Gamma = \Om h$. Lower values of $\Gamma$ lead to larger flows at fixed $\sigma_8$ and $f(\Om)$. There is some cancellation of the $\Om$-dependent terms, and so we have found that the flows depend on the combination $\Om h^2$.  Fig. \ref{fig:omegamhhsig} shows the 2D parameter space of $\Om h^2$ and $\sigma_8$ with $h$, $\Omega_b$ and $n_s$ are held fixed at their WMAP5 central values for this calculation.  The large scale flow found here favours the corner with low $\Om h^2$  and high $\sigma_8$. Also plotted are the WMAP5 68\% and 95\% confidence regions from \cite{DunKomNol08}, as well as the WMAP5+BAO+SN combination from \cite{KomDunNol08}. One can see that the entire WMAP5 95\% CL parameter space is excluded at better than 95\% CL ($2\sigma$).   The WMAP5 ``best case'' (BC) parameters, i.e. those that lie within the \cite{DunKomNol08} 95\% CL regions, have  $\Om h^2 = 0.136$ and $\sigma_8=0.863$ but the r.m.s. flow differs only slightly from the central WMAP5 value, and hence the $\chi^2$ value is very similar (see Table \ref{tab:chisq}).  Essentially both WMAP5 and flows are sensitive to the same large scales, and so they have the same parameter degeneracies.

Another approach we took is maximum-likelihood analysis. The cosmic variance in the bulk flow is a function of the cosmological parameters, and so one can ask which parameters maximize the probability of generating a flow equal to that observed. Basically, given a set of power spectrum parameters ${\bf\Theta}$, the likelihood is given by
\begin{equation}
{\cal L}({\bf\Theta}) =  {1\over |R|^{1/2}}\exp\left(\sum_{ij}-{1\over 2}u_iR^{-1}_{ij}u_j\right).
\label{eq:likelihood}
\end{equation}
where $i$ and $j$ both go from 1 to 3 to specify the bulk flow components.    Taking the natural logarithm of the above, and multiplying by $-2$, we see that this is very similar to the ``frequentist'' $\chi^2$ statistic (Eq. \ref{eq:chi}) except for a term $\ln |R|$ that penalizes models with high cosmic variance.

\begin{figure}
\includegraphics[width=\columnwidth]{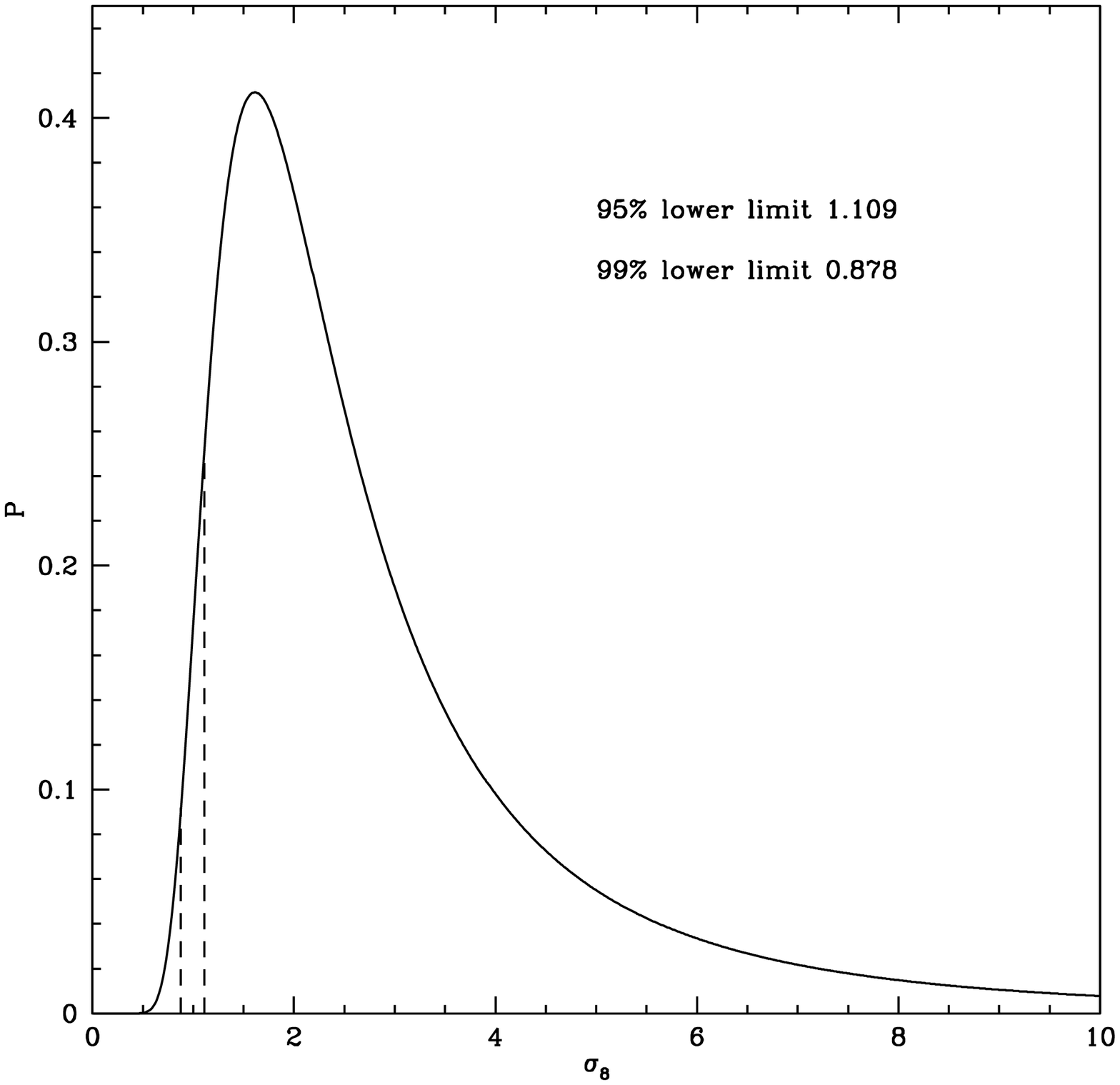}
\caption{Likelihood of the value of $\sigma_8$ when the power spectrum shape is fixed at the WMAP5 central parameters. The dashed lines indicate the 99\% and 95\% lower limits.
}
\label{fig:siglike}
\end{figure}

From Figure \ref{fig:omegamhhsig} we see that the bulk flow is more weakly dependent on $\Om h^2$ than $\sigma_8$.   This can be understood by considering the fact that increasing $\Om$ increases the $f(\Om)$,  but also increases $\Gamma$; these two changes act on the bulk flow in opposite directions and tend to cancel out.  Given the stronger dependence on $\sigma_8$, we have chosen to plot the likelihood of $\sigma_8$ with other cosmological parameters fixed at the central WMAP5 values.   The results are show in Fig. \ref{fig:siglike}. The peak of the likelihood is at $\sigma_8 \sim 1.7$, but the likelihood is very non-Gaussian: it has a sharp edge at low $\sigma_8$ and has a tail extending to very large values. The 1-sided 95\% and 99\% lower limits are 1.109 and 0.878, respectively. Of course these limits would be reduced if one placed priors on $\sigma_8$ that excluded very large values.

\section{Discussion}
\label{sec:discuss}

There are several possible explanations for the discrepancy between observations and model that we have observed. 

First, it is possible that the large observed flow is the result of a systematic error in the data. However, the independence of the distance indicators (TF, FP and SN Ia) and methodology of the various surveys, as well as the agreement between different surveys makes this quite unlikely.  Indeed, examination of Table~\ref{tab:bfv} shows that no one survey is ``pulling'' the COMPOSITE bulk flow.  Furthermore, we have shown that, on the $R_I = 50 \hmpc$ scale, the surveys are mutually consistent within their random errors. Thus systematic errors affecting individual surveys must be small, and, since most systematic errors are expected to be independent, their net effect is smaller still. The only systematic that is likely to affect all of these surveys in the same way is a coherent (dipole-like) error in the foreground Galactic extinctions \citep{SchFinDav98} used to correct magnitudes, and hence distances and peculiar velocities.  This possibility, however, has been tightly constrained via extragalactic ``colour'' standards \citep{Hud99}. 

Second, there is the rather un-Copernican possibility that we happen to live in a rare local volume that has a statistically unlikely large bulk flow magnitude.  The rareness would then be at the 1-in-100 level.

Finally, there is the possibility that the WMAP5-calibrated cosmological model underestimates the amplitude of large-scale fluctuations in the low-$z$ Universe.  In this context, it is interesting to compare our result to those of other independent low-redshift probes, which we consider from small to large scales.

There are a number of small-scale measurements of $\sigma_8$ (or $\Om^{0.5}\sigma_8$) from a variety of techniques, some of which are summarized in \cite{BonConPen05}. The most recent weak-lensing result from the CFHTLS \citep{FuSemHoe08} find  $\sigma_8(\Om / 0.25)^{0.64} = 0.785 \pm 0.043$ when all angular scales are analyzed, but $\sigma_8(\Om / 0.25)^{0.53} = 0.837 \pm 0.084$ when only scales in the linear regime are studied. Some recent results that suggest higher values of $\sigma_8$ include the Lyman-$\alpha$ forest study of \cite{SelSloMcD06}, who find that that the power is enhanced over the expectation of WMAP3, and the study of \cite{ReiAdeBoc08} which finds $\sigma_8 = 0.94^{+0.03}_{-0.04}$ from secondary cluster SZ anisotropies in the CMB. Using pairwise velocity statistics \cite{pairwise03} found $\sigma_8=1.13^{+0.22}_{-0.23}$. 

It is possible to use peculiar velocities to probe matter density fluctuations on scales smaller than those probed by the bulk flow by comparing density-field predictions with observed peculiar velocities point-by-point. As discussed above, this yields an estimate of $\beta \equiv f(\Om)/b$, which, when combined with an independent estimate of $\sigma_{8, gal}$, can be used to determine $f(\Om)\sigma_8$. \cite{PikHud05} combined their analysis of the 2MASS density field with previous comparisons based on IRAS density fields.  They found the degenerate combination $\sigma_8\,(\Om/0.25)^{0.55} = 0.88\pm0.05$.  The quoted error is likely a slight underestimate, since not all studies are independent. Nevertheless, we note that the central value is somewhat higher than the WMAP5 value, although still lower than our 95\% lower limit.  Although the \cite{PikHud05} measurement is based on peculiar velocities, the scale probed is quite different --- in such studies, the large-scale bulk flow is essentially subtracted out, and so the scale is typically that of superclusters: $\sim 20 \hmpc$. Similar statistical results based on peculiar velocities can be obtained from redshift-space distortions; for example, from SDSS LRGs, for $\Om = 0.245$, \cite{CabGaz08} quote $\sigma_8 = 0.85\pm0.06$, in good agreement with \cite{PikHud05}, but still in conflict with our formal 95\% lower limit on $\sigma_8$.  In summary, from data on small to intermediate scales, there are hints that $\sigma_8$ may be higher than the WMAP5 best-fit value, but only by a modest amount which is not sufficient to comfortably explain the large-scale bulk flow.

The bulk flow result result found here suggests significant power on scales of  $k < 0.02$ h/Mpc. While the CMB probes such scales at very high redshift, there is only one other probe of the \emph{matter} power spectrum at low redshift: the ISW effect. This probe shows a signal that is  \emph{stronger} \citep{GazManMul06}, by a factor of $2.2\pm0.6$ \citep{HoHirPad08} than expected from the clustering of galaxies and quasars. It is interesting that our most-likely normalization $\sigma_8$ is also a factor $\sim 2$ larger than the standard model. 

Finally, we consider the result of \cite{KasAtrKoc08b,KasAtrKoc08a}. As noted by those authors, taken at face value, their flow amplitude and, more importantly, the scale over which this flow is observed ($\sim 600 \hmpc$) is greatly in excess of that expected in $\Lambda$CDM models. In this case, it seems impossible to generate cosmologically-consisent results simply by tinkering with the parameters of $\Lambda$CDM; instead a wholesale revision of the model would be called for.

\section{Conclusions}
\label{sec:conc}

We have calculated optimal ``minimum variance'' weights designed to measure bulk flows with minimal sensitivity to small scale power, and have applied these weights to a number of recent peculiar velocity surveys.  We find that all of the surveys we studied are consistent with each other, with the possible exception of the \cite{LauPos94} BCG survey. Taken together the data suggest that the bulk flow within a Gaussian window of radius 50 \hmpc\ is $407\pm81$ \kms\ toward  \lb{287\arcdeg\pm9}{8\arcdeg\pm6}.  This motion is not due to nearby sources, such as the Great Attractor (at a distance of $\sim40$\hmpc), but rather to sources at greater depths that have yet to be fully identified. 

A flow of this amplitude on such a large scale is not expected in the WMAP5-normalized $\Lambda$CDM cosmology. The observed bulk flow favors the upper values of the WMAP5 $\Om h^2$-$\sigma_8$ error-ellipse, but even the point at the top of the WMAP5 95\% confidence ellipse predicts a bulk flow which is too small compared to that observed at a confidence level $> 98$\%.

There are several possible explanations for the discrepancy we have observed.   There is the possibility that we happen to live in a volume with a statistically unlikely ($\lesssim 2$\%) bulk flow magnitude.   If this is the case, then the structures that cause this flow should be eventually identified as the depth and sky coverage of redshift surveys increase.    Alternatively, it is possible that the large observed flow is the result of a systematic error in the data, although the independence of the distance indicators (TF, FP and SN Ia) and methodology of the various surveys, as well as the agreement between different surveys makes this unlikely. 

The bulk flow in the nearby ($d \lesssim 60 \hmpc$) Universe is no longer noise-limited but rather cosmic variance limited, so that increasing the quantity of nearby peculiar velocity data will not alter the significance of this result. At very large depths ($d > 100 \hmpc$), however, the bulk flow measurement is still quite noisy.  Future peculiar velocity surveys, such as the NOAO Fundamental Plane Survey \citep{SmiHudNel04}, as well as nearby supernovae surveys \citep{FilLiTre01,WooAldLee04,KelSchBes07, FriBasBec08}, are expected to yield more precise measurements of the amplitude of the bulk motion on these very large scales, and thus have the potential to strengthen the cosmological constraints therefrom. In order to measure the bulk flow variance directly, one must measure the bulk flow in independent (i.e. distant) regions. For the standard distance estimators used in traditional peculiar velocity work, the errors grow proportional to distance and hence become infeasible at large distances. So other techniques, such as those based on the kinetic Sunyaev-Zel'dovich effect \citep{SunZel72,RepLah91,HaeTeg96, RuhAdeCar04, Kos06, ZhaFelJus08, KasAtrKoc08a} will be needed to access independent volumes.

To reiterate, the results presented in this paper pose a challenge to the standard $\Lambda$CDM model with the WMAP5 parameters. As this study shows, the implications to the standard scenario should be explored with as many independent cosmological observations as we can muster, with particular attention paid to clues from probes at low redshift and on the largest scales.

\vspace{0.2cm}

\noindent{\bf Acknowlegements:} 
We thank Niayesh Afshordi and Russell Smith for interesting discussions and helpful comments on earlier versions of this paper. HAF would like to thank Danny Marfatia for helpful comments and Brad Klee for programming help. HAF has been supported in part by a grant from the Research Corporation, by an NSF grant AST-0807326 and by the National Science Foundation through TeraGrid resources provided by the NCSA. MJH is supported by NSERC. 

\bibliographystyle{mn2e}
\bibliography{mjh}

\begin{thebibliography}{86}
\expandafter\ifx\csname natexlab\endcsname\relax\def\natexlab#1{#1}\fi

\bibitem[{{Aaronson} {et~al.}(1989){Aaronson}, {Bothun}, {Cornell}, {Dawe},
  {Dickens}, {Hall}, {Sheng}, {Huchra}, {Lucey}, {Mould}, {Murray}, {Schommer},
  \& {Wright}}]{AarBotCor89}
{Aaronson} M., {Bothun} G.~D., {Cornell} M.~E., {Dawe} J.~A., {Dickens} R.~J.,
  {Hall} P.~J., {Sheng} H.~M., {Huchra} J.~P., {Lucey} J.~R., {Mould} J.~R.,
  {Murray} J.~D., {Schommer} R.~A., {Wright} A.~E., 1989, \apj, 338, 654

\bibitem[{{Abate} \& {Lahav}(2008)}]{AbaLah08}
{Abate} A., {Lahav} O., 2008, MNRAS Lett, 389, L47

\bibitem[{{Bernardi} {et~al.}(2002){Bernardi}, {Alonso}, {da Costa}, {Willmer},
  {Wegner}, {Pellegrini}, {Rit{\' e}}, \& {Maia}}]{BerAlodaC02b}
{Bernardi} M., {Alonso} M.~V., {da Costa} L.~N., {Willmer} C.~N.~A., {Wegner}
  G., {Pellegrini} P.~S., {Rit{\' e}} C., {Maia} M.~A.~G., 2002, \aj, 123, 2990

\bibitem[{{Bond} {et~al.}(2005){Bond}, {Contaldi}, {Pen}, {Pogosyan}, {Prunet},
  {Ruetalo}, {Wadsley}, {Zhang}, {Mason}, {Myers}, {Pearson}, {Readhead},
  {Sievers}, \& {Udomprasert}}]{BonConPen05}
{Bond} J.~R., {Contaldi} C.~R., {Pen} U.-L., {Pogosyan} D., {Prunet} S.,
  {Ruetalo} M.~I., {Wadsley} J.~W., {Zhang} P., {Mason} B.~S., {Myers} S.~T.,
  {Pearson} T.~J., {Readhead} A.~C.~S., {Sievers} J.~L., {Udomprasert} P.~S.,
  2005, \apj, 626, 12

\bibitem[{{Branchini} {et~al.}(1999){Branchini}, {Teodoro}, {Frenk},
  {Schmoldt}, {Efstathiou}, {White}, {Saunders}, {Sutherland},
  {Rowan-Robinson}, {Keeble}, {Tadros}, {Maddox}, \& {Oliver}}]{BraTeoFre99}
{Branchini} E., {Teodoro} L., {Frenk} C.~S., {Schmoldt} I., {Efstathiou} G.,
  {White} S.~D.~M., {Saunders} W., {Sutherland} W., {Rowan-Robinson} M.,
  {Keeble} O., {Tadros} H., {Maddox} S., {Oliver} S., 1999, \mnras, 308, 1

\bibitem[{{Cabre} \& {Gazta\~naga}(2008)}]{CabGaz08}
{Cabre} A., {Gazta\~naga} E., 2008, ArXiv e-prints, 0807.2460

\bibitem[{{Clutton-Brock} \& {Peebles}(1981)}]{CluPee81}
{Clutton-Brock} M., {Peebles} P.~J.~E., 1981, \aj, 86, 1115

\bibitem[{{Colless} {et~al.}(2001){Colless}, {Saglia}, {Burstein}, {Davies},
  {McMahan}, \& {Wegner}}]{ColSagBur01}
{Colless} M., {Saglia} R.~P., {Burstein} D., {Davies} R.~L., {McMahan} R.~K.,
  {Wegner} G., 2001, \mnras, 321, 277

\bibitem[{{Cooray} \& {Caldwell}(2006)}]{CooCal06}
{Cooray} A., {Caldwell} R.~R., 2006, \prd, 73, 103002

\bibitem[{{Courteau}(1992)}]{Cou92}
{Courteau} S., 1992, \pasp, 104, 976

\bibitem[{{Courteau} {et~al.}(2000){Courteau}, {Willick}, {Strauss},
  {Schlegel}, \& {Postman}}]{CouWilStr00}
{Courteau} S., {Willick} J.~A., {Strauss} M.~A., {Schlegel} D., {Postman} M.,
  2000, \apj, 544, 636

\bibitem[{{da Costa} {et~al.}(2000){da Costa}, {Bernardi}, {Alonso}, {Wegner},
  {Willmer}, {Pellegrini}, {Rit{\' e}}, \& {Maia}}]{daCBerAlo00}
{da Costa} L.~N., {Bernardi} M., {Alonso} M.~V., {Wegner} G., {Willmer}
  C.~N.~A., {Pellegrini} P.~S., {Rit{\' e}} C., {Maia} M.~A.~G., 2000, \aj,
  120, 95

\bibitem[{{Dale} {et~al.}(1999a){Dale}, {Giovanelli}, {Haynes}, {Campusano}, \&
  {Hardy}}]{DalGioHay99}
{Dale} D.~A., {Giovanelli} R., {Haynes} M.~P., {Campusano} L.~E., {Hardy} E.,
  1999a, \aj, 118, 1489

\bibitem[{{Dale} {et~al.}(1999b){Dale}, {Giovanelli}, {Haynes}, {Campusano},
  {Hardy}, \& {Borgani}}]{DalGioHay99b}
{Dale} D.~A., {Giovanelli} R., {Haynes} M.~P., {Campusano} L.~E., {Hardy} E.,
  {Borgani} S., 1999b, \apjl, 510, L11

\bibitem[{{Dressler} {et~al.}(1987){Dressler}, {Faber}, {Burstein}, {Davies},
  {Lynden-Bell}, {Terlevich}, \& {Wegner}}]{DreFabBur87}
{Dressler} A., {Faber} S.~M., {Burstein} D., {Davies} R.~L., {Lynden-Bell} D.,
  {Terlevich} R.~J., {Wegner} G., 1987, \apjl, 313, L37

\bibitem[{{Dunkley} {et~al.}(2008){Dunkley}, {Komatsu}, {Nolta}, {Spergel},
  {Larson}, {Hinshaw}, {Page}, {Bennett}, {Gold}, {Jarosik}, {Weiland},
  {Halpern}, {Hill}, {Kogut}, {Limon}, {Meyer}, {Tucker}, {Wollack}, \&
  {Wright}}]{DunKomNol08}
{Dunkley} J., {Komatsu} E., {Nolta} M.~R., {Spergel} D.~N., {Larson} D.,
  {Hinshaw} G., {Page} L., {Bennett} C.~L., {Gold} B., {Jarosik} N., {Weiland}
  J.~L., {Halpern} M., {Hill} R.~S., {Kogut} A., {Limon} M., {Meyer} S.~S.,
  {Tucker} G.~S., {Wollack} E., {Wright} E.~L., 2008, ArXiv e-prints, 803

\bibitem[{{Eisenstein} \& {Hu}(1998)}]{EisHu98}
{Eisenstein} D.~J., {Hu} W., 1998, \apj, 496, 605

\bibitem[{{Feldman} {et~al.}(2003){Feldman}, {Juszkiewicz}, {Ferreira},
  {Davis}, {Gazta\~naga}, {Fry}, {Jaffe}, {Chambers}, {da Costa}, {Bernardi},
  {Giovanelli}, {Haynes}, \& {Wegner}}]{pairwise03}
{Feldman} H., {Juszkiewicz} R., {Ferreira} P., {Davis} M., {Gazta\~naga} E.,
  {Fry} J., {Jaffe} A., {Chambers} S., {da Costa} L., {Bernardi} M.,
  {Giovanelli} R., {Haynes} M., {Wegner} G., 2003, \apjl, 596, L131

\bibitem[{{Feldman} {et~al.}(2001){Feldman}, {Frieman}, {Fry}, \&
  {Scoccimarro}}]{FelFriFry01}
{Feldman} H.~A., {Frieman} J., {Fry} J.~N., {Scoccimarro} R., 2001, {PRL}, 86,
  1434

\bibitem[{{Feldman} \& {Watkins}(2008)}]{FelWat08}
{Feldman} H.~A., {Watkins} R., 2008, \mnras, 387, 825

\bibitem[{{Filippenko} {et~al.}(2001){Filippenko}, {Li}, {Treffers}, \&
  {Modjaz}}]{FilLiTre01}
{Filippenko} A.~V., {Li} W.~D., {Treffers} R.~R., {Modjaz} M., 2001, in
  Astronomical Society of the Pacific Conference Series, Vol. 246, IAU Colloq.
  183: Small Telescope Astronomy on Global Scales, {Paczynski} B., {Chen}
  W.-P., {Lemme} C., eds., pp. 121--+

\bibitem[{{Fosalba} \& {Dor{\'e}}(2007)}]{FosDor07}
{Fosalba} P., {Dor{\'e}} O., 2007, \prd, 76, 103523

\bibitem[{{Frieman} {et~al.}(2008){Frieman}, {Bassett}, {Becker}, {Choi},
  {Cinabro}, {DeJongh}, {Depoy}, {Dilday}, {Doi}, {Garnavich}, {Hogan},
  {Holtzman}, {Im}, {Jha}, {Kessler}, {Konishi}, {Lampeitl}, {Marriner},
  {Marshall}, {McGinnis}, {Miknaitis}, {Nichol}, {Prieto}, {Riess}, {Richmond},
  {Romani}, {Sako}, {Schneider}, {Smith}, {Takanashi}, {Tokita}, {van der
  Heyden}, {Yasuda}, {Zheng}, {Adelman-McCarthy}, {Annis}, {Assef},
  {Barentine}, {Bender}, {Blandford}, {Boroski}, {Bremer}, {Brewington},
  {Collins}, {Crotts}, {Dembicky}, {Eastman}, {Edge}, {Edmondson}, {Elson},
  {Eyler}, {Filippenko}, {Foley}, {Frank}, {Goobar}, {Gueth}, {Gunn},
  {Harvanek}, {Hopp}, {Ihara}, {Ivezi{\'c}}, {Kahn}, {Kaplan}, {Kent},
  {Ketzeback}, {Kleinman}, {Kollatschny}, {Kron}, {Krzesi{\'n}ski}, {Lamenti},
  {Leloudas}, {Lin}, {Long}, {Lucey}, {Lupton}, {Malanushenko}, {Malanushenko},
  {McMillan}, {Mendez}, {Morgan}, {Morokuma}, {Nitta}, {Ostman}, {Pan},
  {Rockosi}, {Romer}, {Ruiz-Lapuente}, {Saurage}, {Schlesinger}, {Snedden},
  {Sollerman}, {Stoughton}, {Stritzinger}, {Subba Rao}, {Tucker}, {Vaisanen},
  {Watson}, {Watters}, {Wheeler}, {Yanny}, \& {York}}]{FriBasBec08}
{Frieman} J.~A., {Bassett} B., {Becker} A., {Choi} C., {Cinabro} D., {DeJongh}
  F., {Depoy} D.~L., {Dilday} B., {Doi} M., {Garnavich} P.~M., {Hogan} C.~J.,
  {Holtzman} J., {Im} M., {Jha} S., {Kessler} R., {Konishi} K., {Lampeitl} H.,
  {Marriner} J., {Marshall} J.~L., {McGinnis} D., {Miknaitis} G., {Nichol}
  R.~C., {Prieto} J.~L., {Riess} A.~G., {Richmond} M.~W., {Romani} R., {Sako}
  M., {Schneider} D.~P., {Smith} M., {Takanashi} N., {Tokita} K., {van der
  Heyden} K., {Yasuda} N., {Zheng} C., {Adelman-McCarthy} J., {Annis} J.,
  {Assef} R.~J., {Barentine} J., {Bender} R., {Blandford} R.~D., {Boroski}
  W.~N., {Bremer} M., {Brewington} H., {Collins} C.~A., {Crotts} A., {Dembicky}
  J., {Eastman} J., {Edge} A., {Edmondson} E., {Elson} E., {Eyler} M.~E.,
  {Filippenko} A.~V., {Foley} R.~J., {Frank} S., {Goobar} A., {Gueth} T.,
  {Gunn} J.~E., {Harvanek} M., {Hopp} U., {Ihara} Y., {Ivezi{\'c}} {\v Z}.,
  {Kahn} S., {Kaplan} J., {Kent} S., {Ketzeback} W., {Kleinman} S.~J.,
  {Kollatschny} W., {Kron} R.~G., {Krzesi{\'n}ski} J., {Lamenti} D., {Leloudas}
  G., {Lin} H., {Long} D.~C., {Lucey} J., {Lupton} R.~H., {Malanushenko} E.,
  {Malanushenko} V., {McMillan} R.~J., {Mendez} J., {Morgan} C.~W., {Morokuma}
  T., {Nitta} A., {Ostman} L., {Pan} K., {Rockosi} C.~M., {Romer} A.~K.,
  {Ruiz-Lapuente} P., {Saurage} G., {Schlesinger} K., {Snedden} S.~A.,
  {Sollerman} J., {Stoughton} C., {Stritzinger} M., {Subba Rao} M., {Tucker}
  D., {Vaisanen} P., {Watson} L.~C., {Watters} S., {Wheeler} J.~C., {Yanny} B.,
  {York} D., 2008, \aj, 135, 338

\bibitem[{{Fu} {et~al.}(2008){Fu}, {Semboloni}, {Hoekstra}, {Kilbinger}, {van
  Waerbeke}, {Tereno}, {Mellier}, {Heymans}, {Coupon}, {Benabed}, {Benjamin},
  {Bertin}, {Dor{\'e}}, {Hudson}, {Ilbert}, {Maoli}, {Marmo}, {McCracken}, \&
  {M{\'e}nard}}]{FuSemHoe08}
{Fu} L., {Semboloni} E., {Hoekstra} H., {Kilbinger} M., {van Waerbeke} L.,
  {Tereno} I., {Mellier} Y., {Heymans} C., {Coupon} J., {Benabed} K.,
  {Benjamin} J., {Bertin} E., {Dor{\'e}} O., {Hudson} M.~J., {Ilbert} O.,
  {Maoli} R., {Marmo} C., {McCracken} H.~J., {M{\'e}nard} B., 2008, \aap, 479,
  9

\bibitem[{{Gazta{\~n}aga} {et~al.}(2006){Gazta{\~n}aga}, {Manera}, \&
  {Multam{\"a}ki}}]{GazManMul06}
{Gazta{\~n}aga} E., {Manera} M., {Multam{\"a}ki} T., 2006, \mnras, 365, 171

\bibitem[{{Giovanelli} {et~al.}(1998a){Giovanelli}, {Haynes}, {Freudling}, {da
  Costa}, {Salzer}, \& {Wegner}}]{GioHayFre98}
{Giovanelli} R., {Haynes} M.~P., {Freudling} W., {da Costa} L.~N., {Salzer}
  J.~J., {Wegner} G., 1998a, \apjl, 505, L91

\bibitem[{{Giovanelli} {et~al.}(1998b){Giovanelli}, {Haynes}, {Salzer},
  {Wegner}, {da Costa}, \& {Freudling}}]{GioHaySal98}
{Giovanelli} R., {Haynes} M.~P., {Salzer} J.~J., {Wegner} G., {da Costa} L.~N.,
  {Freudling} W., 1998b, \aj, 116, 2632

\bibitem[{{Haehnelt} \& {Tegmark}(1996)}]{HaeTeg96}
{Haehnelt} M.~G., {Tegmark} M., 1996, \mnras, 279, 545

\bibitem[{{Han} \& {Mould}(1992)}]{HanMou92}
{Han} M., {Mould} J.~R., 1992, \apj, 396, 453

\bibitem[{{Haugboelle} {et~al.}(2007){Haugboelle}, {Hannestad}, {Thomsen},
  {Fynbo}, {Sollerman}, \& {Jha}}]{HauHanTho06}
{Haugboelle} T., {Hannestad} S., {Thomsen} B., {Fynbo} J., {Sollerman} J.,
  {Jha} S., 2007, ApJ, 661, 650

\bibitem[{{Ho} {et~al.}(2008){Ho}, {Hirata}, {Padmanabhan}, {Seljak}, \&
  {Bahcall}}]{HoHirPad08}
{Ho} S., {Hirata} C., {Padmanabhan} N., {Seljak} U., {Bahcall} N., 2008, \prd,
  78, 043519

\bibitem[{{Hoffman} {et~al.}(2001){Hoffman}, {Eldar}, {Zaroubi}, \&
  {Dekel}}]{HofEldZar01}
{Hoffman} Y., {Eldar} A., {Zaroubi} S., {Dekel} A., 2001, ArXiv Astrophysics
  e-prints

\bibitem[{{Hudson}(1994{\natexlab{a}})}]{Hud94}
{Hudson} M.~J., 1994{\natexlab{a}}, \mnras, 266, 468

\bibitem[{{Hudson}(1994{\natexlab{b}})}]{Hud94b}
---, 1994{\natexlab{b}}, \mnras, 266, 475

\bibitem[{{Hudson}(1999)}]{Hud99}
---, 1999, \pasp, 111, 57

\bibitem[{Hudson(2003)}]{Hud03}
Hudson M.~J., 2003, in Proceedings of the 15th Rencontres De Blois: Physical
  Cosmology: New Results In Cosmology And The Coherence Of The Standard Model,
  Bartlett J., ed., in press (astro-ph preprint 0311072)

\bibitem[{{Hudson} \& {Ebeling}(1997)}]{HudEbe97}
{Hudson} M.~J., {Ebeling} H., 1997, \apj, 479, 621

\bibitem[{{Hudson} {et~al.}(1997){Hudson}, {Lucey}, {Smith}, \&
  {Steel}}]{HudLucSmi97}
{Hudson} M.~J., {Lucey} J.~R., {Smith} R.~J., {Steel} J., 1997, \mnras, 291,
  488

\bibitem[{{Hudson} {et~al.}(2004){Hudson}, {Smith}, {Lucey}, \&
  {Branchini}}]{HudSmiLuc04}
{Hudson} M.~J., {Smith} R.~J., {Lucey} J.~R., {Branchini} E., 2004, \mnras,
  352, 61

\bibitem[{{Hudson} {et~al.}(1999){Hudson}, {Smith}, {Lucey}, {Schlegel}, \&
  {Davies}}]{HudSmiLuc99}
{Hudson} M.~J., {Smith} R.~J., {Lucey} J.~R., {Schlegel} D.~J., {Davies} R.~L.,
  1999, \apjl, 512, L79

\bibitem[{{Hui} \& {Greene}(2006)}]{HuiGre06}
{Hui} L., {Greene} P.~B., 2006, \prd, 73, 123526

\bibitem[{{Juszkiewicz} {et~al.}(2000){Juszkiewicz}, {Ferreira}, {Feldman},
  {Jaffe}, \& {Davis}}]{pairwise00}
{Juszkiewicz} R., {Ferreira} P.~G., {Feldman} H.~A., {Jaffe} A.~H., {Davis} M.,
  2000, Science, 287, 109

\bibitem[{{Kaiser}(1988)}]{Kai88}
{Kaiser} N., 1988, \mnras, 231, 149

\bibitem[{{Kashlinsky} {et~al.}(2008{\natexlab{a}}){Kashlinsky},
  {Atrio-Barandela}, {Kocevski}, \& {Ebeling}}]{KasAtrKoc08b}
{Kashlinsky} A., {Atrio-Barandela} F., {Kocevski} D., {Ebeling} H.,
  2008{\natexlab{a}}, \apjl, 686, L49

\bibitem[{{Kashlinsky} {et~al.}(2008{\natexlab{b}}){Kashlinsky},
  {Atrio-Barandela}, {Kocevski}, \& {Ebeling}}]{KasAtrKoc08a}
---, 2008{\natexlab{b}}, ArXiv e-prints

\bibitem[{{Keller} {et~al.}(2007){Keller}, {Schmidt}, {Bessell}, {Conroy},
  {Francis}, {Granlund}, {Kowald}, {Oates}, {Martin-Jones}, {Preston},
  {Tisserand}, {Vaccarella}, \& {Waterson}}]{KelSchBes07}
{Keller} S.~C., {Schmidt} B.~P., {Bessell} M.~S., {Conroy} P.~G., {Francis} P.,
  {Granlund} A., {Kowald} E., {Oates} A.~P., {Martin-Jones} T., {Preston} T.,
  {Tisserand} P., {Vaccarella} A., {Waterson} M.~F., 2007, Publications of the
  Astronomical Society of Australia, 24, 1

\bibitem[{{Kocevski} {et~al.}(2004){Kocevski}, {Mullis}, \&
  {Ebeling}}]{KocEbeMul04}
{Kocevski} D.~D., {Mullis} C.~R., {Ebeling} H., 2004, \apj, 608, 721

\bibitem[{{Komatsu} {et~al.}(2008){Komatsu}, {Dunkley}, {Nolta}, {Bennett},
  {Gold}, {Hinshaw}, {Jarosik}, {Larson}, {Limon}, {Page}, {Spergel},
  {Halpern}, {Hill}, {Kogut}, {Meyer}, {Tucker}, {Weiland}, {Wollack}, \&
  {Wright}}]{KomDunNol08}
{Komatsu} E., {Dunkley} J., {Nolta} M.~R., {Bennett} C.~L., {Gold} B.,
  {Hinshaw} G., {Jarosik} N., {Larson} D., {Limon} M., {Page} L., {Spergel}
  D.~N., {Halpern} M., {Hill} R.~S., {Kogut} A., {Meyer} S.~S., {Tucker} G.~S.,
  {Weiland} J.~L., {Wollack} E., {Wright} E.~L., 2008, ArXiv e-prints, 803

\bibitem[{{Kosowsky}(2006)}]{Kos06}
{Kosowsky} A., 2006, New Astronomy Review, 50, 969

\bibitem[{{Lauer} \& {Postman}(1994)}]{LauPos94}
{Lauer} T.~R., {Postman} M., 1994, \apj, 425, 418

\bibitem[{Linder(2005)}]{Lin05}
Linder E.~V., 2005, PRD, 72, 043529

\bibitem[{{Lynden-Bell} {et~al.}(1988){Lynden-Bell}, {Faber}, {Burstein},
  {Davies}, {Dressler}, {Terlevich}, \& {Wegner}}]{LynFabBur88}
{Lynden-Bell} D., {Faber} S.~M., {Burstein} D., {Davies} R.~L., {Dressler} A.,
  {Terlevich} R.~J., {Wegner} G., 1988, \apj, 326, 19

\bibitem[{{Mathewson} {et~al.}(1992){Mathewson}, {Ford}, \&
  {Buchhorn}}]{MatForBuc92b}
{Mathewson} D.~S., {Ford} V.~L., {Buchhorn} M., 1992, \apjs, 81, 413

\bibitem[{{Neill} {et~al.}(2007){Neill}, {Hudson}, \& {Conley}}]{NeiHudCon07}
{Neill} J.~D., {Hudson} M.~J., {Conley} A., 2007, \apjl, 661, L123

\bibitem[{{Peebles}(1993)}]{PPC}
{Peebles} P. J.~E., 1993, Principles of Physical Cosmology. Princeton
  University Press

\bibitem[{{Pike} \& {Hudson}(2005)}]{PikHud05}
{Pike} R.~W., {Hudson} M.~J., 2005, \apj, 635, 11

\bibitem[{{Postman} \& {Lauer}(1995)}]{PosLau95}
{Postman} M., {Lauer} T.~R., 1995, \apj, 440, 28

\bibitem[{{Reichardt} {et~al.}(2008){Reichardt}, {Ade}, {Bock}, {Bond},
  {Brevik}, {Contaldi}, {Daub}, {Dempsey}, {Goldstein}, {Holzapfel}, {Kuo},
  {Lange}, {Lueker}, {Newcomb}, {Peterson}, {Ruhl}, {Runyan}, \&
  {Staniszewski}}]{ReiAdeBoc08}
{Reichardt} C.~L., {Ade} P.~A.~R., {Bock} J.~J., {Bond} J.~R., {Brevik} J.~A.,
  {Contaldi} C.~R., {Daub} M.~D., {Dempsey} J.~T., {Goldstein} J.~H.,
  {Holzapfel} W.~L., {Kuo} C.~L., {Lange} A.~E., {Lueker} M., {Newcomb} M.,
  {Peterson} J.~B., {Ruhl} J., {Runyan} M.~C., {Staniszewski} Z., 2008, ArXiv
  e-prints, 801

\bibitem[{{Rephaeli} \& {Lahav}(1991)}]{RepLah91}
{Rephaeli} Y., {Lahav} O., 1991, \apj, 372, 21

\bibitem[{{Rowan-Robinson} {et~al.}(2000){Rowan-Robinson}, {Sharpe}, {Oliver},
  {Keeble}, {Canavezes}, {Saunders}, {Taylor}, {Valentine}, {Frenk},
  {Efstathiou}, {McMahon}, {White}, {Sutherland}, {Tadros}, \&
  {Maddox}}]{RowShaOli00}
{Rowan-Robinson} M., {Sharpe} J., {Oliver} S.~J., {Keeble} O., {Canavezes} A.,
  {Saunders} W., {Taylor} A.~N., {Valentine} H., {Frenk} C.~S., {Efstathiou}
  G.~P., {McMahon} R.~G., {White} S.~D.~M., {Sutherland} W., {Tadros} H.,
  {Maddox} S., 2000, \mnras, 314, 375

\bibitem[{{Rubin} {et~al.}(1976){Rubin}, {Roberts}, {Thonnard}, \&
  {Ford}}]{RubRobTho76}
{Rubin} V.~C., {Roberts} M.~S., {Thonnard} N., {Ford} W.~K., 1976, \aj, 81, 719

\bibitem[{{Ruhl} {et~al.}(2004){Ruhl}, {Ade}, {Carlstrom}, {Cho}, {Crawford},
  {Dobbs}, {Greer}, {Halverson}, {Holzapfel}, {Lanting}, {Lee}, {Leitch},
  {Leong}, {Lu}, {Lueker}, {Mehl}, {Meyer}, {Mohr}, {Padin}, {Plagge}, {Pryke},
  {Runyan}, {Schwan}, {Sharp}, {Spieler}, {Staniszewski}, \&
  {Stark}}]{RuhAdeCar04}
{Ruhl} J., {Ade} P.~A.~R., {Carlstrom} J.~E., {Cho} H.-M., {Crawford} T.,
  {Dobbs} M., {Greer} C.~H., {Halverson} N.~w., {Holzapfel} W.~L., {Lanting}
  T.~M., {Lee} A.~T., {Leitch} E.~M., {Leong} J., {Lu} W., {Lueker} M., {Mehl}
  J., {Meyer} S.~S., {Mohr} J.~J., {Padin} S., {Plagge} T., {Pryke} C.,
  {Runyan} M.~C., {Schwan} D., {Sharp} M.~K., {Spieler} H., {Staniszewski} Z.,
  {Stark} A.~A., 2004, in Presented at the Society of Photo-Optical
  Instrumentation Engineers (SPIE) Conference, Vol. 5498, Millimeter and
  Submillimeter Detectors for Astronomy II. Edited by Jonas Zmuidzinas, Wayne
  S. Holland and Stafford Withington Proceedings of the SPIE, Volume 5498, pp.
  11-29 (2004)., {Bradford} C.~M., {Ade} P.~A.~R., {Aguirre} J.~E., {Bock}
  J.~J., {Dragovan} M., {Duband} L., {Earle} L., {Glenn} J., {Matsuhara} H.,
  {Naylor} B.~J., {Nguyen} H.~T., {Yun} M., {Zmuidzinas} J., eds., pp. 11--29

\bibitem[{{Sachs} \& {Wolfe}(1967)}]{SacWol67}
{Sachs} R.~K., {Wolfe} A.~M., 1967, \apj, 147, 73

\bibitem[{{Sarkar} {et~al.}(2007){Sarkar}, {Feldman}, \&
  {Watkins}}]{SarFelWat07}
{Sarkar} D., {Feldman} H.~A., {Watkins} R., 2007, \mnras, 375, 691

\bibitem[{{Schlegel} {et~al.}(1998){Schlegel}, {Finkbeiner}, \&
  {Davis}}]{SchFinDav98}
{Schlegel} D.~J., {Finkbeiner} D.~P., {Davis} M., 1998, \apj, 500, 525

\bibitem[{{Scoccimarro} {et~al.}(2001){Scoccimarro}, {Feldman}, {Frieman}, \&
  {Fry}}]{ScoFelFri01}
{Scoccimarro} R., {Feldman} H.~A., {Frieman} J., {Fry} J.~N., 2001, \apj, 546,
  652

\bibitem[{{Seljak} {et~al.}(2006){Seljak}, {Slosar}, \&
  {McDonald}}]{SelSloMcD06}
{Seljak} U., {Slosar} A., {McDonald} P., 2006, Journal of Cosmology and
  Astro-Particle Physics, 10, 14

\bibitem[{{Smith} {et~al.}(2004){Smith}, {Hudson}, {Nelan}, {Moore}, {Quinney},
  {Wegner}, {Lucey}, {Davies}, {Malecki}, {Schade}, \&
  {Suntzeff}}]{SmiHudNel04}
{Smith} R.~J., {Hudson} M.~J., {Nelan} J.~E., {Moore} S.~A.~W., {Quinney}
  S.~J., {Wegner} G.~A., {Lucey} J.~R., {Davies} R.~L., {Malecki} J.~J.,
  {Schade} D., {Suntzeff} N.~B., 2004, \aj, 128, 1558

\bibitem[{{Springob} {et~al.}(2007){Springob}, {Masters}, {Haynes},
  {Giovanelli}, \& {Marinoni}}]{SprMasHay08}
{Springob} C.~M., {Masters} K.~L., {Haynes} M.~P., {Giovanelli} R., {Marinoni}
  C., 2007, \apjs, 172, 599

\bibitem[{{Strauss} {et~al.}(1992){Strauss}, {Yahil}, {Davis}, {Huchra}, \&
  {Fisher}}]{StrYahDav92}
{Strauss} M.~A., {Yahil} A., {Davis} M., {Huchra} J.~P., {Fisher} K., 1992,
  \apj, 397, 395

\bibitem[{{Sunyaev} \& {Zeldovich}(1972)}]{SunZel72}
{Sunyaev} R.~A., {Zeldovich} Y.~B., 1972, Comments on Astrophysics and Space
  Physics, 4, 173

\bibitem[{{Tonry} {et~al.}(2001){Tonry}, {Dressler}, {Blakeslee}, {Ajhar},
  {Fletcher}, {Luppino}, {Metzger}, \& {Moore}}]{TonDreBla01}
{Tonry} J.~L., {Dressler} A., {Blakeslee} J.~P., {Ajhar} E.~A., {Fletcher}
  A.~B., {Luppino} G.~A., {Metzger} M.~R., {Moore} C.~B., 2001, \apj, 546, 681

\bibitem[{{Tonry} {et~al.}(2003){Tonry}, {Schmidt}, {Barris}, {Candia},
  {Challis}, {Clocchiatti}, {Coil}, {Filippenko}, {Garnavich}, {Hogan},
  {Holland}, {Jha}, {Kirshner}, {Krisciunas}, {Leibundgut}, {Li}, {Matheson},
  {Phillips}, {Riess}, {Schommer}, {Smith}, {Sollerman}, {Spyromilio},
  {Stubbs}, \& {Suntzeff}}]{TonSchBar03}
{Tonry} J.~L., {Schmidt} B.~P., {Barris} B., {Candia} P., {Challis} P.,
  {Clocchiatti} A., {Coil} A.~L., {Filippenko} A.~V., {Garnavich} P., {Hogan}
  C., {Holland} S.~T., {Jha} S., {Kirshner} R.~P., {Krisciunas} K.,
  {Leibundgut} B., {Li} W., {Matheson} T., {Phillips} M.~M., {Riess} A.~G.,
  {Schommer} R., {Smith} R.~C., {Sollerman} J., {Spyromilio} J., {Stubbs}
  C.~W., {Suntzeff} N.~B., 2003, \apj, 594, 1

\bibitem[{{Tully} {et~al.}(2008){Tully}, {Shaya}, {Karachentsev}, {Courtois},
  {Kocevski}, {Rizzi}, \& {Peel}}]{TulShaKar08}
{Tully} R.~B., {Shaya} E.~J., {Karachentsev} I.~D., {Courtois} H.~M.,
  {Kocevski} D.~D., {Rizzi} L., {Peel} A., 2008, \apj, 676, 184

\bibitem[{{Verde} {et~al.}(2002){Verde}, {Heavens}, {Percival}, {Matarrese},
  {Baugh}, {Bland-Hawthorn}, {Bridges}, {Cannon}, {Cole}, {Colless}, {Collins},
  {Couch}, {Dalton}, {De Propris}, {Driver}, {Efstathiou}, {Ellis}, {Frenk},
  {Glazebrook}, {Jackson}, {Lahav}, {Lewis}, {Lumsden}, {Maddox}, {Madgwick},
  {Norberg}, {Peacock}, {Peterson}, {Sutherland}, \& {Taylor}}]{VerHeaPer02}
{Verde} L., {Heavens} A.~F., {Percival} W.~J., {Matarrese} S., {Baugh} C.~M.,
  {Bland-Hawthorn} J., {Bridges} T., {Cannon} R., {Cole} S., {Colless} M.,
  {Collins} C., {Couch} W., {Dalton} G., {De Propris} R., {Driver} S.~P.,
  {Efstathiou} G., {Ellis} R.~S., {Frenk} C.~S., {Glazebrook} K., {Jackson} C.,
  {Lahav} O., {Lewis} I., {Lumsden} S., {Maddox} S., {Madgwick} D., {Norberg}
  P., {Peacock} J.~A., {Peterson} B.~A., {Sutherland} W., {Taylor} K., 2002,
  \mnras, 335, 432

\bibitem[{{Vittorio} {et~al.}(1986){Vittorio}, {Juszkiewicz}, \&
  {Davis}}]{VitJusDav86}
{Vittorio} N., {Juszkiewicz} R., {Davis} M., 1986, \nat, 323, 132

\bibitem[{{Watkins} \& {Feldman}(1995)}]{WatFel95}
{Watkins} R., {Feldman} H.~A., 1995, \apjl, 453, L73

\bibitem[{{Watkins} \& {Feldman}(2007)}]{WatFel07}
---, 2007, \mnras, 379, 343

\bibitem[{{Wegner} {et~al.}(2003){Wegner}, {Bernardi}, {Willmer}, {da Costa},
  {Alonso}, {Pellegrini}, {Maia}, {Chaves}, \& {Rit{\'e}}}]{WegBerWil03}
{Wegner} G., {Bernardi} M., {Willmer} C.~N.~A., {da Costa} L.~N., {Alonso}
  M.~V., {Pellegrini} P.~S., {Maia} M.~A.~G., {Chaves} O.~L., {Rit{\'e}} C.,
  2003, \aj, 126, 2268

\bibitem[{{Willick}(1990)}]{Wil90}
{Willick} J.~A., 1990, \apjl, 351, L5

\bibitem[{{Willick}(1999)}]{Wil99b}
---, 1999, \apj, 522, 647

\bibitem[{{Willick} {et~al.}(1997){Willick}, {Courteau}, {Faber}, {Burstein},
  {Dekel}, \& {Strauss}}]{WilCouFab97}
{Willick} J.~A., {Courteau} S., {Faber} S.~M., {Burstein} D., {Dekel} A.,
  {Strauss} M.~A., 1997, \apjs, 109, 333

\bibitem[{{Wood-Vasey} {et~al.}(2004){Wood-Vasey}, {Aldering}, {Lee}, {Loken},
  {Nugent}, {Perlmutter}, {Siegrist}, {Wang}, {Antilogus}, {Astier}, {Hardin},
  {Pain}, {Copin}, {SMadja}, {Gangler}, {Castera}, {Adam}, {Bacon},
  {Lemonnier}, {Pecontal}, {Pecontal}, \& {Kessler}}]{WooAldLee04}
{Wood-Vasey} W.~M., {Aldering} G., {Lee} B.~C., {Loken} S., {Nugent} P.,
  {Perlmutter} S., {Siegrist} J., {Wang} L., {Antilogus} P., {Astier} P.,
  {Hardin} D., {Pain} R., {Copin} Y., {SMadja} G., {Gangler} E., {Castera} A.,
  {Adam} G., {Bacon} R., {Lemonnier} J.-P., {Pecontal} A., {Pecontal} E.,
  {Kessler} R., 2004, New Astron.Rev, 48, 637

\bibitem[{{Zaroubi}(2002)}]{Zar02}
{Zaroubi} S., 2002, \mnras, 331, 901

\bibitem[{{Zaroubi} {et~al.}(1999){Zaroubi}, {Hoffman}, \&
  {Dekel}}]{ZarHofDek99}
{Zaroubi} S., {Hoffman} Y., {Dekel} A., 1999, \apj, 520, 413

\bibitem[{{Zhang} {et~al.}(2008){Zhang}, {Feldman}, {Juszkiewicz}, \&
  {Stebbins}}]{ZhaFelJus08}
{Zhang} P., {Feldman} H.~A., {Juszkiewicz} R., {Stebbins} A., 2008, \mnras,
  388, 735

\end{thebibliography}

\end{document}